\documentclass[preprintnumbers,amsmath,amssymb]{revtex4}
\usepackage{graphicx}
\usepackage{dcolumn}% Align table columns on decimal point
\usepackage{bm}% bold math
\begin{document}
\def\oppropto{\mathop{\propto}} 
\def\opsimeq{\mathop{\simeq}}
\def\opoverderline{\mathop{\overline}}
\def\operarrow{\mathop{\longrightarrow}}
\def\opsim{\mathop{\sim}} 

\title{Numerical study of the disordered Poland-Scheraga model of DNA denaturation}
\author{Thomas Garel and C\'ecile Monthus}
 \affiliation{Service de Physique Th\'{e}orique, CEA/DSM/SPhT\\
Unit\'e de recherche associ\'ee au CNRS\\
91191 Gif-sur-Yvette cedex, France}

\begin{abstract}
We numerically study the binary disordered Poland-Scheraga model of DNA
denaturation, in the regime where the pure model displays a first
order transition (loop exponent $c=2.15>2$). We use a Fixman-Freire
scheme for the entropy of loops and consider chain length up to  
$N=4 \cdot 10^5$, with averages over $10^4$ samples.
We present in parallel the results of various observables
 for two boundary conditions,
namely bound-bound (bb) and bound-unbound (bu), because they present very
different finite-size behaviors, both in the pure case and
in the disordered case. Our main conclusion is that the transition
remains first order in the disordered case: in the (bu) case, the 
disorder averaged energy and contact densities present crossings for
different values of $N$ without rescaling. In addition, we obtain that
these disorder averaged observables do not satisfy finite size scaling, as a
consequence of strong sample to sample fluctuations of the
pseudo-critical temperature. For a given 
sample, we propose a procedure to identify its pseudo-critical
temperature, and show that this sample then obeys first order
transition finite size scaling behavior. Finally, we obtain that the 
disorder averaged critical loop distribution is still governed by
$P(l) \sim 1/l^c$ in the regime $l \ll N$, as in the pure case.

\bigskip

%PACS numbers: 87.14.Gg; 87.15.Cc; 82.39.Pj

\end{abstract}
\maketitle

\section{Introduction}

Recent Monte Carlo simulations of three dimensional interacting self
avoiding walks (SAW's) \cite{Barbara1, Carlon, Baiesi1, Baiesi2} have
revived interest \cite{Ka_Mu_Pe1,Ka_Mu_Pe2,Rich_Gutt,Schaf} in the
Poland-Scheraga (PS) model of DNA denaturation \cite{Pol_Scher}. In
this model, the central parameter is the exponent $c$ governing the
weight $1/l^c$ of a loop of length $l$. If the binding energy is
constant along the chain, the 
model is exactly soluble, and the transition is first order for $c>2$
and second order for $1<c<2$. Gaussian loops in $d$ dimensions are
characterized by $c=c_G=\frac{d}{2}$. The role of self avoidance
within a loop was taken into account by Fisher \cite{Fisher}, and
yields $c=d\nu$, where $\nu$ is the SAW radius of gyration
exponent. More recently, Kafri et al. \cite{Ka_Mu_Pe1,Ka_Mu_Pe2}
pointed out that the inclusion of the self avoidance of the loop with
the rest of the chains further increased $c$ to a value $c>2$, both in
$d=2$ and $d=3$ (see also \cite{Dup}). The transition in the pure
model should then be discontinuous, as previously found by
\cite{Barbara1} in Monte Carlo simulations of SAW 's. The value $c
\simeq 2.11$ was in turn measured in three dimensional Monte Carlo
simulations by  \cite{Carlon, Baiesi1, Baiesi2}. 

The precise correspondence between the PS model and the numerical
simulations of SAW's in $d$ dimensional space is a matter of debate: the
comparison of the finite-size properties of both models \cite{Schaf}
involves not only the loop weight $1/l^c$ discussed above, but also
the end segment properties, since simulations use SAW's bound at the
origin, but with free end points. The PS model can easily be
formulated for free boundary conditions, but this requires some prescription
for the weight of the unbound free ends. The correct prescription to
mimic the SAWs model is still controversial \cite{Ka_Mu_Pe2,Baiesi1,Schaf}
as explained in more details below.

In this paper, we are interested in the effect of disorder on the PS
model with loop exponent $c=2.15$. This disorder approach is a first
step towards the heterogenity of biological sequences, stemming from
the existence of AT and GC Watson-Crick pairs. We will therefore
consider a binary disorder distribution of the binding energy.
In a physics oriented context, there have been few studies of the
effect of disorder on PS and PS-related models with $c>2$. For SAW's,
Monte-Carlo simulations have been interpreted as follows (i) in
\cite{Carlon}, the loop exponent $c>2$ was found to remain the same as
in the pure case for two particular biological sequences  
(ii) the more complete study of the binary disordered case in ref
\cite{Barbara2} suggests a smoothening out of the transition, without
fully excluding the possibility of a first order transition.

On the theoretical side, there exists many discussions
on the effect of disorder on first-order transitions
\cite{Im_Wo,Ai_We,Hu_Be,Cardy} but these studies typically consider
spin systems displaying coexisting domains in the pure case,
so that the conclusions of these studies cannot be directly applied to
a polymer transition, such as the present PS transition.
In this paper, we study numerically the disordered PS model
for a binary distribution of the binding energies.
We use the same method as in our recent study \cite{Gar_Mon} of the disordered
two-dimensional wetting model (corresponding to the PS model with $c=3/2$).
We present in parallel results
for two different boundary conditions bound-bound (bb) and
bound-unbound (bu), since these two boundary conditions lead to very
different finite size behaviors, as already stressed in \cite{Schaf}
for the pure case.

\section{Poland-Scheraga model of DNA denaturation}

\subsection{Model and observables}

We consider here pairing interactions between monomers on the
two chains, in contrast to the stacking interactions present in the usual
treatment of DNA melting \cite{WB85, Meltsim,Ga_Or}. The
Poland-Scheraga description of DNA denaturation \cite{Pol_Scher} can
be defined in the following way: we consider a forward partition
function $Z_f(\alpha)$  for two chains of $(\alpha)$ monomers, with
monomer pairs $(1)$ and $(\alpha)$ bound together. 

\begin{figure}[htbp]
\begin{center}
\includegraphics{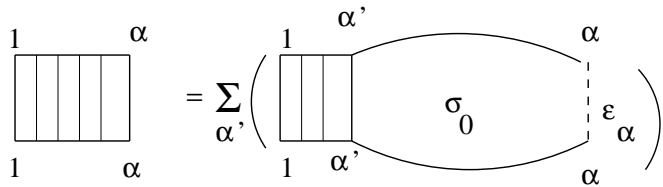}
\end{center}
\caption{Recursion relation for $Z_f(\alpha)$ (eq.(\ref{forward})) in
the PS model with pairing interactions $\varepsilon_{\alpha}$.} 
\label{f0}
\end{figure}

This partition function
obeys the recursion relation (see Fig \ref{f0})
\begin{equation}
\label{forward}
Z_f(\alpha)=e^{-\beta \varepsilon_{\alpha}} \sum_{\alpha ^{\prime
}=1,3,..}^{\alpha -2}Z_f(\alpha ^{\prime })
\mathcal{N}(\alpha ^{\prime };\alpha )
\end{equation}
where $\mathcal{N}(\alpha ^{\prime };\alpha )$ is the the partition
function of a loop going from $\alpha ^{\prime }$ to $\alpha$. In
polymer physics, the
asymptotic expansion of $\mathcal{N}(\alpha ^{\prime };\alpha )$ for
($\alpha-\alpha ^{\prime} \gg 1$) is given by \cite{PGG}
\begin{equation}
\label{asymp}
\mathcal{N}(\alpha ^{\prime };\alpha ) \simeq \sigma_0
 \ \mu^{\alpha-\alpha ^{\prime}}\ f(\alpha-\alpha ^{\prime})
\end{equation}
where $\sigma_0$ is a constant, $\mu$ is the fugacity of a pair of monomers
and $f(x)=\frac{1}{x^c}$ is the weight of a loop of length $x \gg
1$. In the PS model, equation (\ref{asymp}) is assumed to hold at all
loop lengths.
The
exponent $c$ is an input of the model, and we study here the case
$c=2.15$, as explained in the Introduction. 
In a similar way, we
define a backward partition 
function $Z_b(\alpha)$, defined as the partition function of the two
chains starting at monomers $(N)$, with monomers $(\alpha)$ bound
together (monomers $(N)$ being bound or unbound). It obeys the
recursion relation 
\begin{equation}
\label{backward}
Z_b(\alpha )=e^{-\beta \varepsilon _{\alpha}} \left(\sum_{\alpha ^{\prime
}=\alpha+2,\alpha+4,..}^{N}Z_b(\alpha ^{\prime })
\mathcal{N}(\alpha; \alpha ^{\prime })+ \mathcal{M}(\alpha;N)\right)
\end{equation}
where
\begin{equation}
\label{asymp2}
\mathcal{M}(\alpha ;N ) \simeq \sigma_1
 \ \mu^{N-\alpha}\ g(\alpha;N)
\end{equation}
is the weight for the possible unpaired segments of length
$(N-\alpha)$, and where 
$\sigma_1$ is a constant. The first term on
the r.h.s of equation (\ref{backward}) is similar to the one of equation
(\ref{forward}). The second one allows for the boundary condition at
monomer $(N)$. For (bb)
boundary conditions, $\sigma_1 =0$, since monomers $(N)$ are bound
together. For (bu) boundary conditions, $\sigma_1 \ne 0$, and the
function $g(\alpha; N)$ is an additional input of the model. According 
to Kafri et al. \cite{Ka_Mu_Pe2}, one has $g(\alpha; N) \simeq
\frac{1}{(N-\alpha)^{c_1}}$ for $(N-\alpha \ll N)$, with $c_1 \simeq
0.092$ for $d=3$. The precise form of $g(\alpha;N)$ beyond the regime
$(N-\alpha \ll N)$ is still controversial \cite{Baiesi1,Schaf}.
Our choice is this paper is to take $g(\alpha; N) \simeq
\frac{1}{(N-\alpha)^{c_1}}$ and $c_1=0$ for all values of $\alpha$, as
is done in biological use of the PS model. Other values of $c_1$
and/or $g(\alpha; N)$ will be briefly considered in the pure case.

In these notations, the thermodynamic partition function $Z$ 
is given by $Z=Z_b(1)$, and the probability for 
monomers $(\alpha)$ to be bound is
\begin{equation}
\label{proba1}
p(\alpha)=\frac {Z_f(\alpha)Z_b(\alpha) e^{\beta
\varepsilon_{\alpha}}}{Z_b(1)}
\end{equation}
where the factor $e^{\beta \varepsilon_{\alpha}}$ in the numerator
avoids double counting of the contact energy at $\alpha$. We will also 
use the unpairing probability $q(\alpha)=1-p(\alpha)$.
A quantity of primary importance in the DNA context is the fraction of 
paired monomers, or contact density
\begin{equation}
\theta_N(T) = \frac{1}{N} \sum_{\alpha=1}^N p(\alpha)
\label{theta}
\end{equation}
In the pure case, $\theta_N(T)$ is proportional to the energy. Since
this is not true in the disordered case, we also consider the (contact)
energy density
\begin{equation}
e_N(T) = \frac{1}{N} \sum_{\alpha=1}^N \varepsilon_{\alpha} \ p(\alpha)
\label{energy}
\end{equation}

We will also be interested in $P_{\rm
loop}(\alpha,\gamma)$, defined as the probability of having
a loop between bounded monomers at $\alpha$ and $\gamma$ 
\begin{equation}
\label{proba2}
P_{\rm loop}(\alpha,\gamma)=\frac
{Z_f(\alpha)\mathcal{N}(\alpha;\gamma)Z_b(\gamma)}{Z_b(1)} 
\end{equation}
As in \cite{Gar_Mon}, we define the probability measure $M_N(l)$
for the loops existing in a sample of length $N$ as follows: for each
loop length $l$, we sum  over all possible positions
$(\alpha,\gamma=\alpha+l)$ of the loop probability $P_{\rm loop}
(\alpha,\alpha+l)$ 
defined in (eq.(\ref{proba2}))
\begin{eqnarray}
M_N(l) = \sum_{\alpha=1}^{N-l} P_{\rm loop} (\alpha,\alpha+l) 
\label{pnl}
\end{eqnarray}
The normalization of this measure over $l$ corresponds
to the averaged number of loops in a sample of size $N$,
or equivalently to the averaged number $N \theta_N (T)$ 
of contacts (\ref{theta}) between the strands.

\subsection{Numerical implementation}

The above equations show that the numerical calculation of 
the thermodynamic partition function $Z=Z_b(1)$ requires a CPU
time of order $O(N^2)$. The Fixman-Freire method \cite{Fix_Fre}
reduces this CPU time to $O(N)$ by approximating the probability
factor $f(x)$ of equation (\ref{asymp}) by  
\begin{equation}
\label{FF}
f(x)=\frac{1 }{{x}^{c}} \simeq f_{FF}(x)=\sum_{k=1}^I a_k \
e^{-b_k x}  \ \ \ \ {\rm for} \ \ \ 1 \leq x \leq N
\end{equation}
\begin{table}[htbp]
\centerline{
\begin{tabular}{|l|l|l|}   \hline
 k  &     $ \ \ \ \ \ \  a_k $ & $ \ \ \ \ \ \ b_k$ \\ \hline
 1  &    7.4210949671864474          &  2.469385023938319     \\ \hline
 2  &     0.7363845614102907        &       0.9006571079384265  \\ \hline
 3  &   0.08732482111472176         &   0.33060668412882793  \\ \hline
 4  &    9.887211230992947.$10^{-3}$&       0.1196018051115684  \\ \hline
 5  & 1.102823738631329.$10^{-3}$   &   0.0430783130904746  \\ \hline
 6  &   1.2250560616947666.$10^{-4}$&       0.015497408698374987  \\ \hline
 7  & 1.3593126590298345.$10^{-5}$  & 5.573286954347416.$10^{-3}$  \\ \hline
 8  &   1.5077329395704116.$10^{-6}$&     2.0039933412704532.$10^{-3}$  \\ \hline
 9  &  1.6718473281512066.$10^{-7}$ & 7.204015696886597.$10^{-4}$  \\ \hline
 10 &    1.852449196083061.$10^{-8}$&     2.5878821693784586.$10^{-4}$  \\ \hline
 11 & 2.0478548398679403.$10^{-9}$  & 9.275640096786728.$10^{-5}$  \\ \hline
 12 &   2.247364959164715.$10^{-10}$&     3.301049580630032.$10^{-5}$  \\ \hline
 13 & 2.4079053140360544.$10^{-11}$ & 1.1480239104095493.$10^{-5}$  \\ \hline
 14 &   2.369755720744333.$10^{-12}$&     3.693407020570475.$10^{-6}$  \\ \hline
 15 & 1.608846567970496.$10^{-13}$  & 8.820504064582594.$10^{-7}$  \\ \hline
\end{tabular}
}

\caption{The coefficients list of the Fixman-Freire scheme
(eq. (\ref{FF})) used in this 
paper.}
\label{tableFF215}
\end{table}

This method is very well known and widely used in biology,
since the standard program MELTSIM, which yields melting curves of
DNA sequences, uses the Fixman-Freire scheme with $I=14$ terms.
This procedure has been tested on DNA chains of length up to $N=10^6$
base pairs \cite{Meltsim,Yer}.

Since the replacement (\ref{FF}) can be rather surprising
for physicists, we ``justify'' it in a more detailed way. One point of
view is to start with the integral representation of the $\Gamma$ function
\begin{equation}
\frac{1}{l^c } = \ {\rm lim}_{M \to \infty} \frac{1}{\Gamma(c)}
\int_0^{M} dt \ t^{c-1} \ e^{- l t} 
\end{equation}
and to discretize it for $L=Ml$ large as
\begin{equation}
\frac{1}{L^c } = \ {\rm lim}_{m \to \infty} \
\sum_{k=1}^{m} \ a_k \ e^{-b_k L} 
\end{equation}
This shows that a power-law can be
represented by a discrete sum of exponentials if
their number $m$, positions $b_k$ and weights $a_k$ are conveniently chosen.
In the Fixman-Freire procedure, the $2I$ coefficients $(a_k,b_k)$ are
obtained from the fit on $2I$ points $(l_1,..,l_{2I})$ such that
$(\ln l_1, \ln l_2, .. , \ln l_{2I})$ divide into equal intervals
the domain $[\ln l_{min},\ln l_{max}]$ where $l_{min}$ and $l_{max}$
are respectively the minimal and maximal loop lengths that are needed
in the numerical program. The number $2I$ of coefficients is then
chosen to obtain the desired numerical accuracy.
It turns out that the choice ${I}=15$ gives an accuracy
better than $0.3\%$. We have adopted this value throughout this paper,
with the values of the coefficients $(a_k,b_k)$ given in Table
\ref{tableFF215}. The reader is invited to draw the log-log plot of the curve
$f_{FF}(l)= \sum_i a_k e^{-b_k l}$ from $l_{min}=1$ up to 
$l_{max}=8 \cdot 10^5$ (maximum value used in this paper): it turns
out that is indistinguishable from a straight line with slope $-2.15$.

Putting everything together, the model we have numerically
studied is defined by recursion equations
(\ref{forward},\ref{backward}) for the partition functions where (i) the
loop partition function $\mathcal{N}(\alpha; \alpha 
^{\prime })$ has been replaced by its asymptotic expression
(\ref{asymp}), with the Fixman-Freire approximation (eq. \ref{FF})
for $f(x)$ (ii) the end segment partition function
$\mathcal{M}(\alpha; N)$  has been replaced by its asymptotic expression
(\ref{asymp2}), with $g(\alpha;
N)=g(N-\alpha)=\frac{1}{(N-\alpha)^{c_1}}$, with $c_1=0$.
We have taken $\sigma_0=0.29607$ and $\sigma_1 =0.5$ for (bu) boundary 
conditions.

Before we turn to the disordered case, we first discuss
the finite-size properties of the pure model
that will be useful to interpret the results of the disordered case. 

\section{Finite size properties of the pure case}

\subsection{ Contact density }
\begin{figure}[htbp]
%\begin{figure}
\includegraphics[height=6cm]{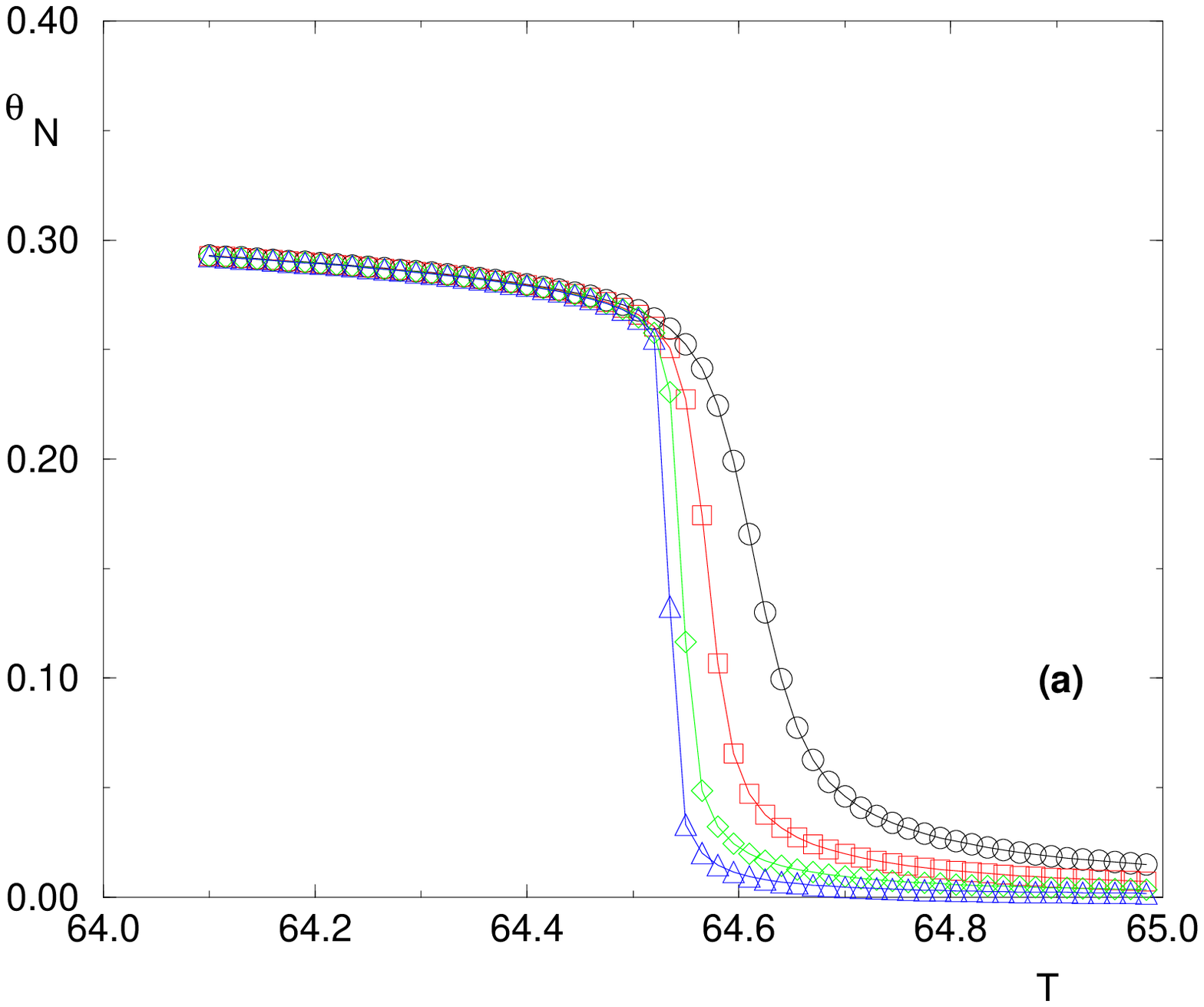}
\hspace{1cm}
\includegraphics[height=6cm]{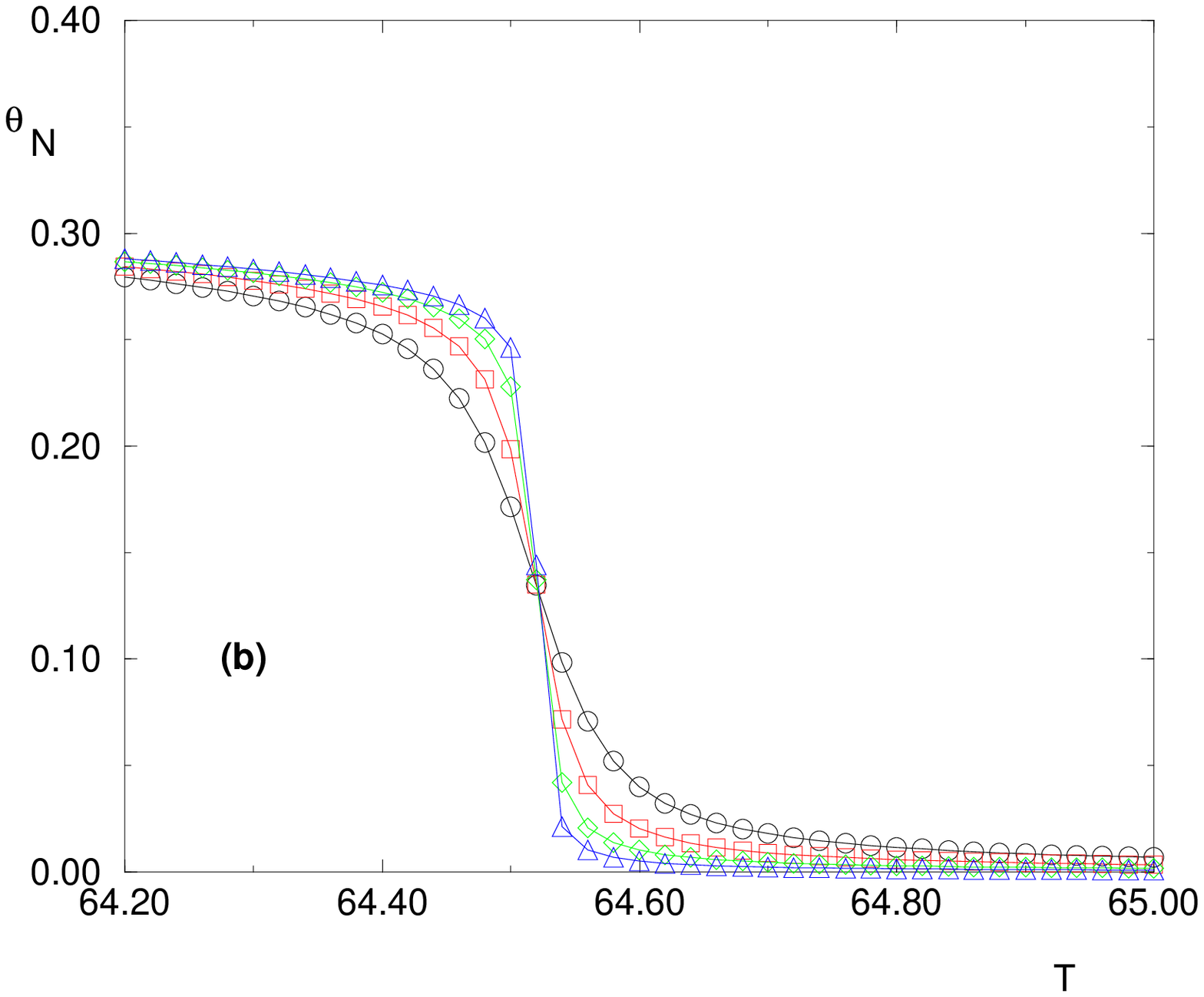}
\caption{The fraction of paired monomers $\theta_N(T)$ of the pure
case  for sizes $N=10^5$ $(\bigcirc)$, $2 \cdot 10^5$ $(\square)$, $4
\cdot 10^5$ $(\diamond)$, $8 \cdot 10^5$ $(\triangle)$. The boundary
conditions are (a) bound-bound (b) bound-unbound.}
\label{f1}
\end{figure}

\begin{figure}[htbp]
%\begin{figure}
\includegraphics[height=6cm]{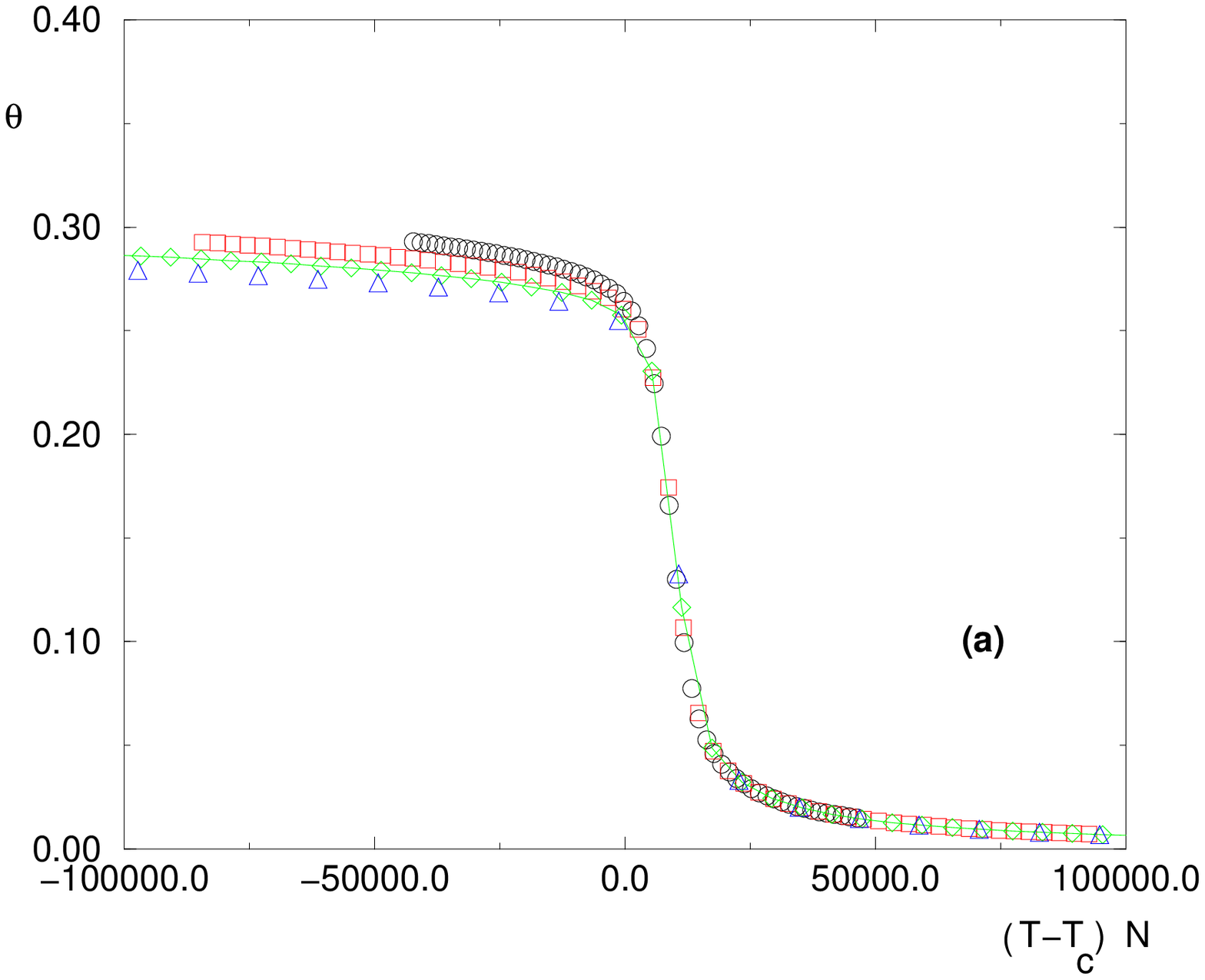}
\hspace{1cm}
\includegraphics[height=6cm]{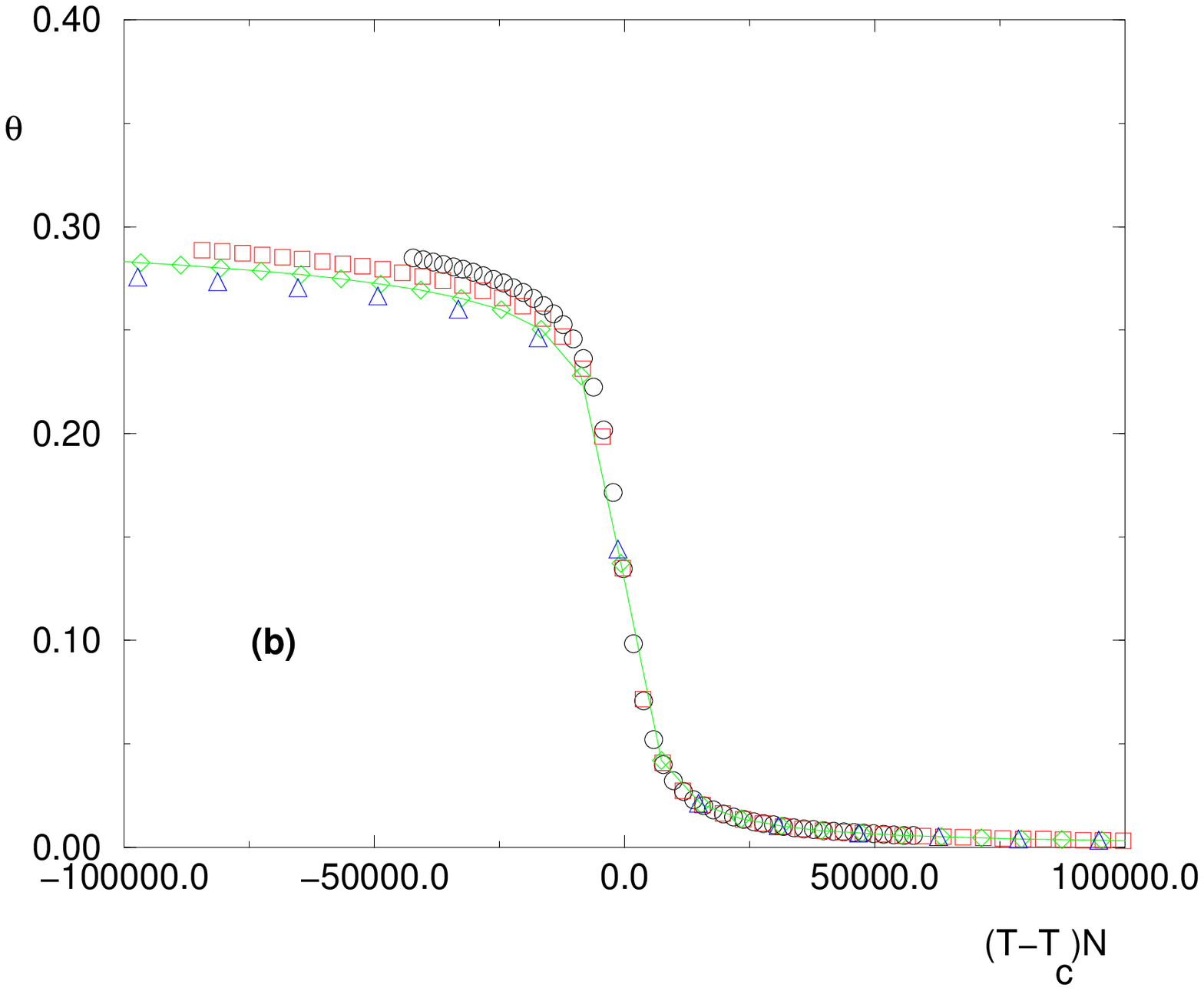}
\caption{The scaled fraction of paired monomers $\theta((T-T_c)N)$ of
the pure case, for $T_c= 64.521..$ for sizes $N=10^5$ $(\bigcirc)$, $2
\cdot 10^5$ $(\square)$, $4 \cdot 10^5$ $(\diamond)$, $8 \cdot 10^5$
$(\triangle)$. The boundary conditions are (a) bound-bound (b)
bound-unbound.} 
\label{f2}
\end{figure}
We choose $\varepsilon_{\alpha}=\varepsilon_0=-355 \ {\rm K}$ (see below).
The contact density $\theta_N(T)$ defined in eq. (\ref{theta})
is plotted on Fig. \ref{f1} for the two boundary conditions
(bound-bound) and (bound-unbound) :
for the (bb) case, there is no crossing at $T_c$ as $N$ varies,
whereas for the (bu) case, there is a crossing at $T_c \simeq
64.521..$  (Celsius scale),
as was also found for the pure SAWs Monte-Carlo simulations
(see Fig. 7 of Ref \cite{Barbara1}).
From a numerical point of view, the boundary conditions (bu) 
seem therefore much more interesting since it yields 
a direct measure of the critical temperature $T_c$.
With the value of $T_c$ obtained from the crossing of Fig. 1 (b),
we may now plot on Fig. 2 the contact density $\theta_N(T)$
in terms of the rescaled variable $(T-T_c)N^{\phi}$
with the crossover exponent $\phi=1$ :
for the two boundary conditions (bb) and (bu),
the finite-size scaling form
\begin{equation}
\theta_N(T) = \Theta( (T-T_c)N ) 
\label{rescalingtheta}
\end{equation}
holds in the delocalized phase $T>T_c$, 
as was also found in the pure SAWs Monte-Carlo simulations
(see Fig. 7 of Ref \cite{Barbara1}).

\subsection{ Specific heat }

\begin{figure}[htbp]
%\begin{figure}
\includegraphics[height=6cm]{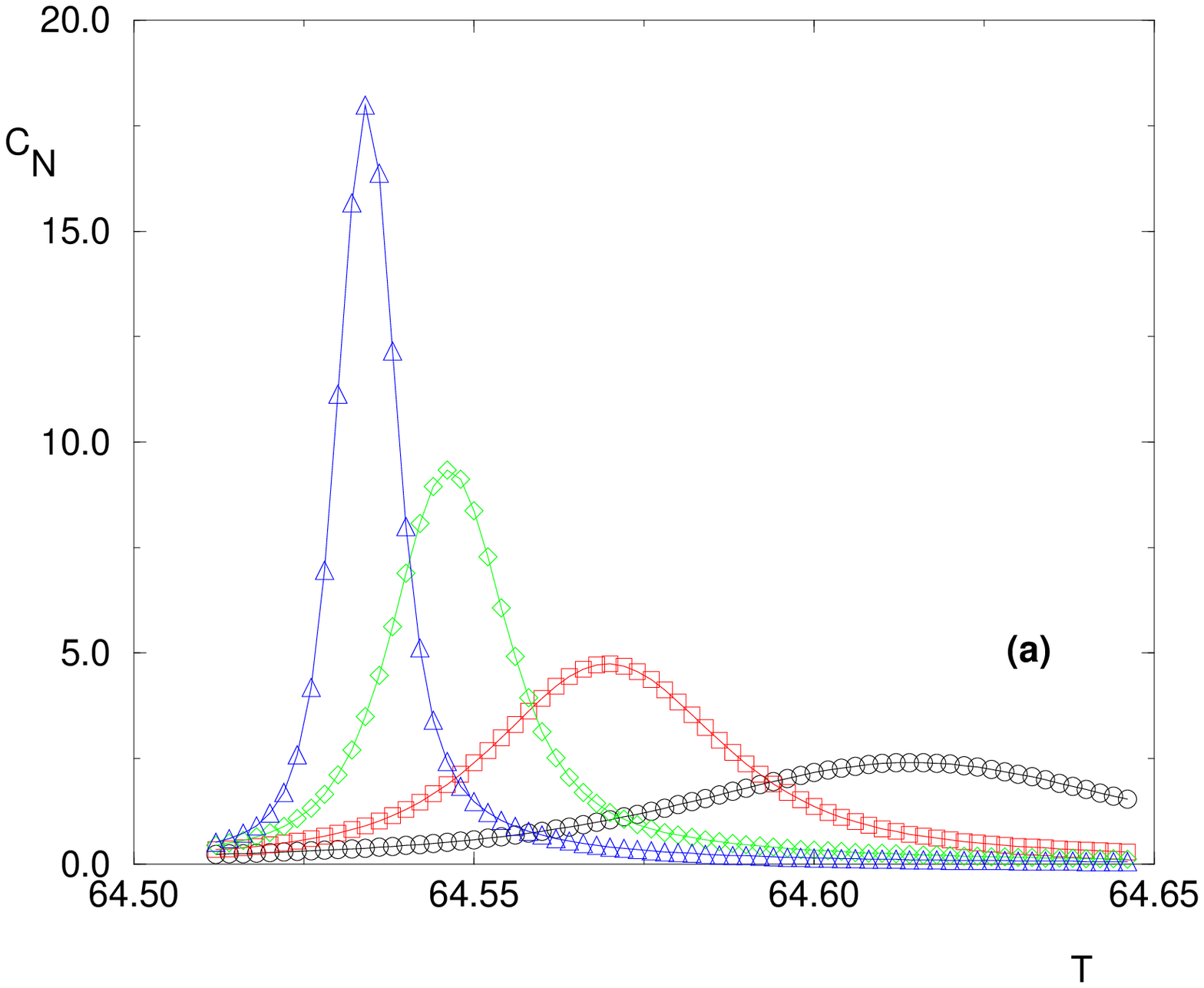}
\hspace{1cm}
\includegraphics[height=6cm]{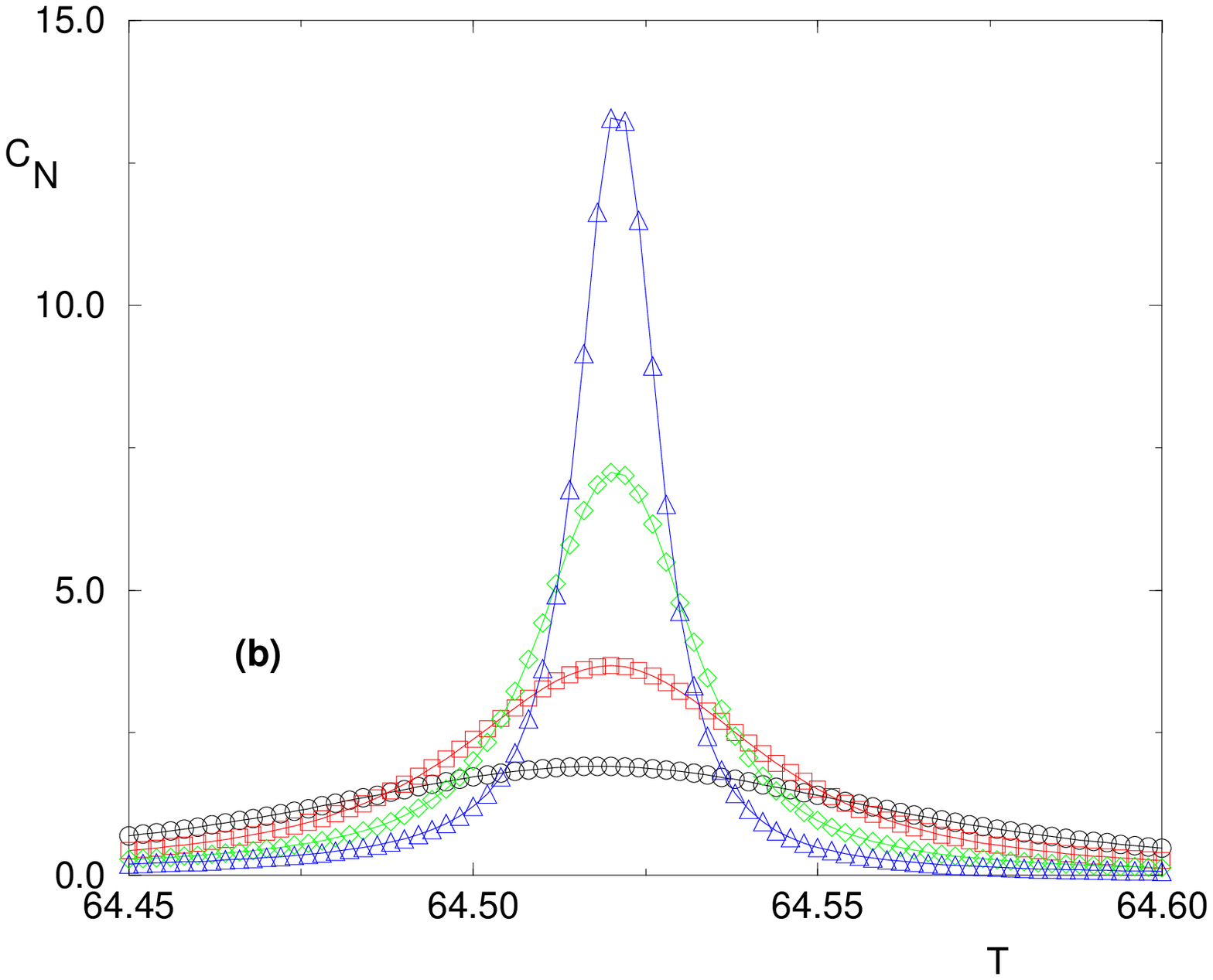}
\caption{The specific heat $C_N$ of the pure case, for
sizes $N=10^5$ $(\bigcirc)$, $2 
\cdot 10^5$ $(\square)$, $4 \cdot 10^5$ $(\diamond)$, $8 \cdot 10^5$
$(\triangle)$. The boundary conditions are (a) bound-bound (b)
bound-unbound.} 
\label{f3}
\end{figure}

The specific heat is shown on Fig. \ref{f3}
for the two boundary conditions :
in both cases, the maximum $C_N^{max}$ scales as the size $N$.
For the (bb) case, the temperature $T_c(N)$ of the maximum 
moves towards the critical temperature $T_c$ as $T_c(N)-T_c \sim 1/N$,  
whereas for the (bu) boundary conditions,
the temperature of the maximum remains almost stable as $N$ varies,
as was also found in the pure SAWs Monte-Carlo simulations
(see Fig. 8 of Ref \cite{Barbara1}).

\subsection{ Pairing probabilities along the chain }

\label{coexistence}

\begin{figure}[htbp]
%\begin{figure}
\includegraphics[height=6cm]{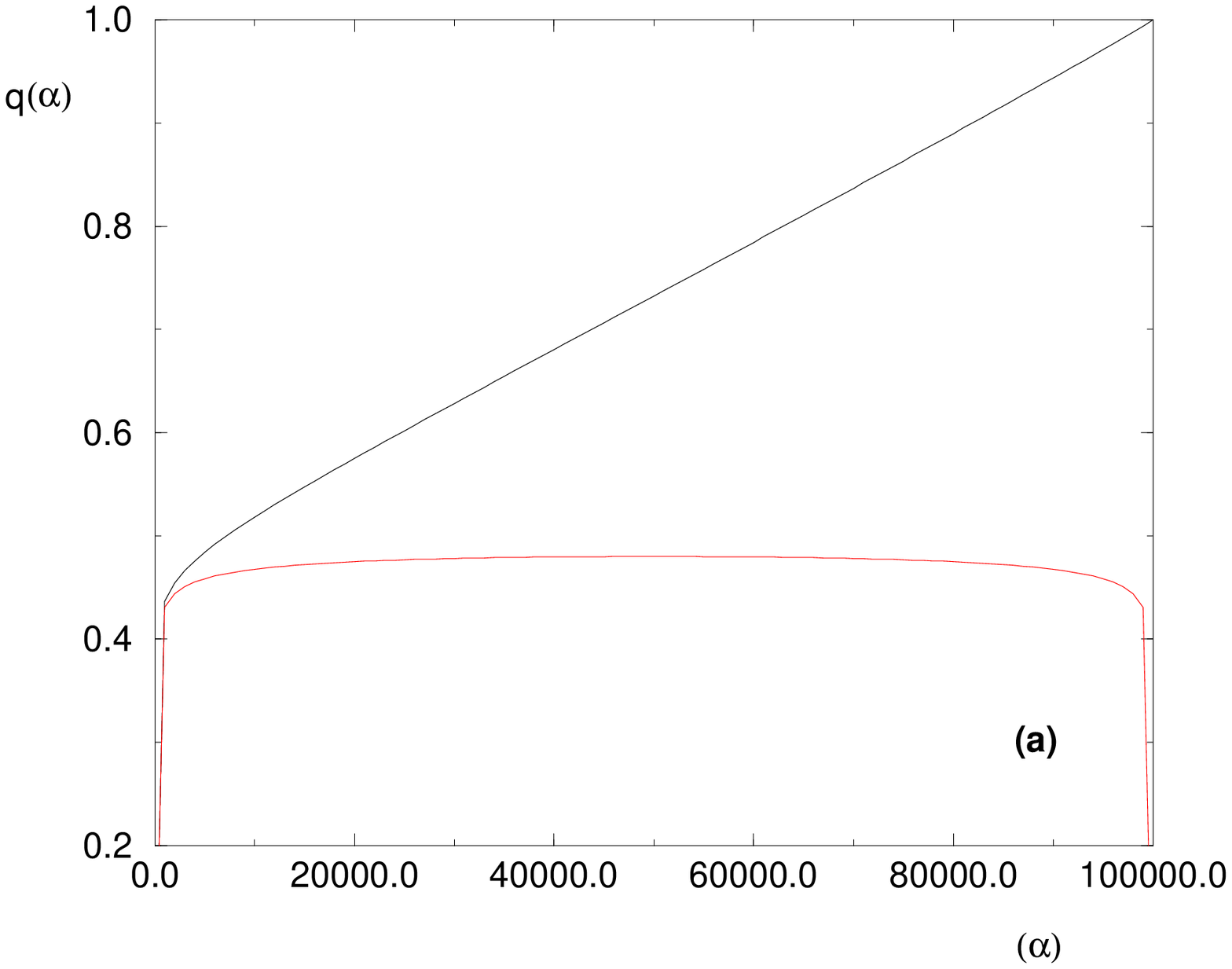}
\hspace{1cm}
\includegraphics[height=6cm]{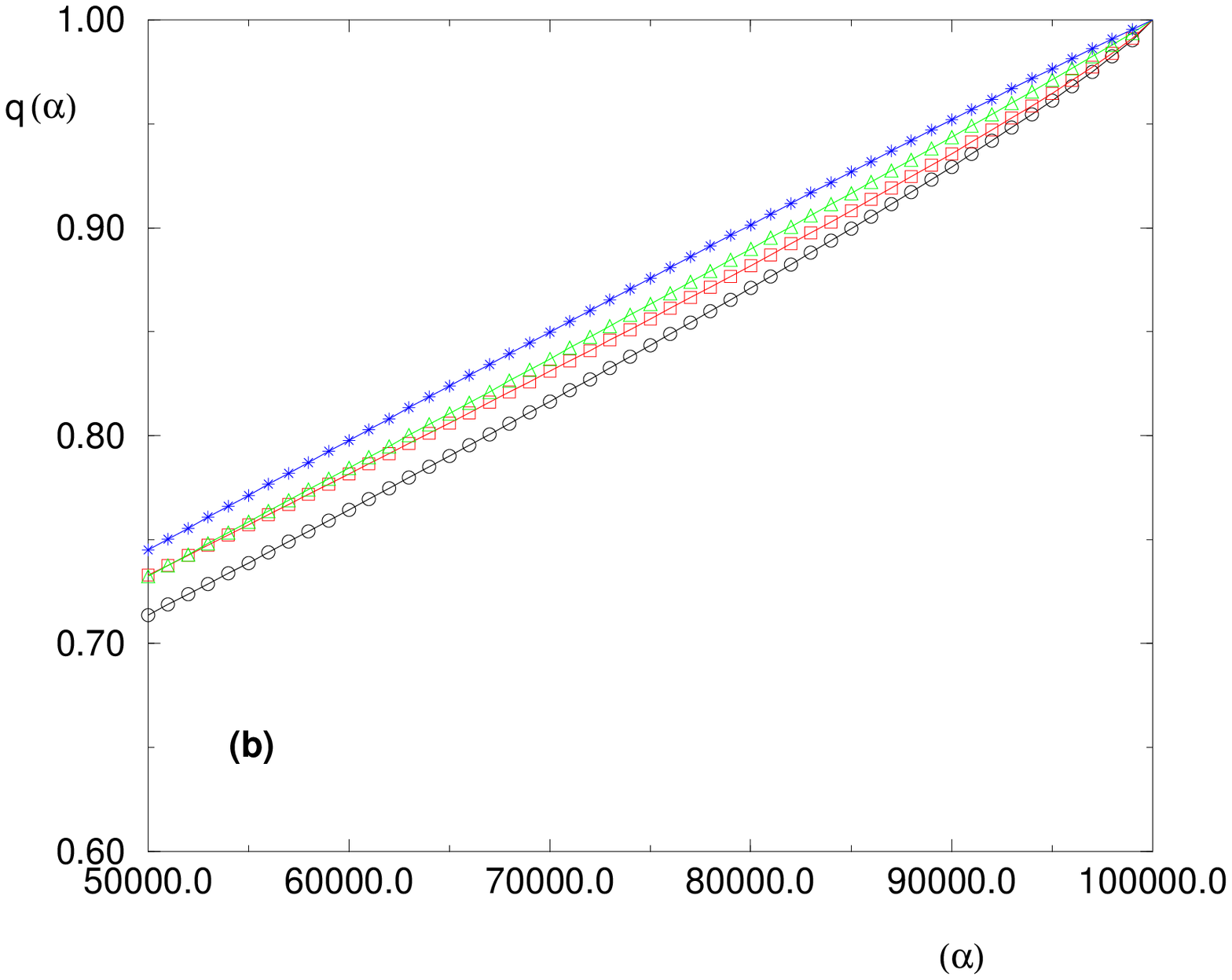}
\caption{(a) The  unpairing probability $q(\alpha)=(1-p(\alpha))$ of
the bound-bound (lower curve) and bound-unbound (upper curve) pure
case at $T_c$, for $N=10^5$ and $c_1=0$ (b) The unpairing 
probability $q(\alpha$), $\alpha > \frac{N}{2}$, of various (bu) pure
cases at $T_c$ for $N=10^5$: $c_1=0$ ($\triangle$), $c_1=0.1$
($\bigcirc$), $c_1=-0.07$ ($\star$), and the prescription of reference
\cite{Schaf} ($\square$). Note that the latter is very
close to the case $c_1=0$ studied in this paper.}
\label{f4}
\end{figure}

On Fig. \ref{f4} (a), we have plotted the unpairing probability 
$q(\alpha)=1-p(\alpha)$ (see eq. \ref{proba1}) as a function of the
monomer index $\alpha$, at criticality, for both boundary conditions : 
for the (bb) boundary condition,
this  unpairing probability is flat around the value $q_c$,
whereas for the (bu) boundary condition,
the  unpairing probability varies linearly as 
\begin{equation}
q(\alpha) \simeq  q_c + (1-q_c) \frac{\alpha}{N}
\label{coexi}
\end{equation}
This linear behavior actually means that there exists a phase coexistence
at $T_c$ between the localized phase on $(1,\alpha_d)$
and the delocalized phase on $(\alpha_d,N)$,
where the position $\alpha_d$ of the interface
is uniformly distributed on $(1,N)$.
In the SAWs simulations, this coexistence 
with a uniform position of the interface
is seen via the flat histogram of the contact density
over Monte-Carlo configurations
(see Fig. 5 of Ref \cite{Barbara1}).
For the (bb) boundary condition, there is no phase coexistence :
the whole sample is in the localized phase.

For the (bu) boundary condition, the presence of a free end
of length $l \sim N$ at criticality means that the choice of the
function $g(\alpha;N)$ in the free end weight of eq. (\ref{asymp2})
may have some importance for phase coexistence properties at
criticality. 

A plausible choice, that we have adopted here, is to extend the equation
$g(\alpha;N)=\frac{1}{(N-\alpha)^{c_1}}$, valid for $(N-\alpha) \ll N$ 
to all values of $\alpha$. We show in Fig \ref{f4} (b) the unpairing
probability $q(\alpha)$ for various values of $c_1$ :
$c_1 \simeq 0.1$  \cite{Ka_Mu_Pe2} and $c_1=0$.

Other choices of the exponent $c_1$ and/or of the function
$g(\alpha;N)$ are possible 
\cite{Baiesi1,Schaf,Schafetal}. We show in particular in Fig \ref{f4}
(b) the unpairing probability $q(\alpha)$ for (i) the value $c_1=-0.07$,
obtained in ref. \cite{Schafetal} for a three arm star polymer (ii) the 
prescription $g(\alpha;N)=
\frac{1}{\alpha^{c_1}}\frac{1}{(N-\alpha)^{c_1}}$, with $c_1 \simeq
0.1$ advocated in ref \cite{Schaf}.

As shown in Fig, \ref{f4}, we obtain only slight deviations with
respect to the straight line behavior of eq. (\ref{coexi}) obtained for our
choice $g(\alpha;N)=\frac{1}{(N-\alpha)^{c_1}}$, $c_1=0$. 

\subsection{ Conclusion on the pure case }

We have shown how the two boundary conditions (bb) and (bu)
lead to different behaviors : at criticality, the (bb) case presents a
pure localized phase, whereas the (bu) case gives rise to a phase
coexistence between a localized phase and a delocalized phase,
with an interface uniformly distributed over the sample.
In this respect, the (bb) case seems simpler. However, from the point
of view of finite size scaling analysis, the (bu) case has the
advantage of allowing a direct measure of the critical temperature via
the crossing of the contact density.

We have also mentioned how the the presents results
for the (bu) case can be qualitatively compared with the pure SAWs simulations
\cite{Barbara1}, and we refer to \cite{Schaf} for 
a detailed quantitative comparison between the two models.

We now turn to the study of the disordered case.

\section{Finite size properties of the disordered case}

We consider the case of binary disorder where the binding energies
$\varepsilon_{\alpha}$ are independent random variables:
$\varepsilon_{\alpha}=\varepsilon_0=-355 \ {\rm K}$
with probability $\frac{1}{2}$ and
$\varepsilon_{\alpha}=\varepsilon_1=-390 \ {\rm K}$ with probability
$\frac{1}{2}$.
As mentioned in the Introduction, this distribution represent AT and GC
base pairs, and the values of ($\varepsilon_0,\varepsilon_1$) are such 
that the critical temperatures of the pure cases are close to the
AT and GC experimental ones. 

Disordered averaged quantities $A$ will be denoted by $\overline{A}$.

\subsection{ Disordered averaged contact and energy densities}

\begin{figure}[htbp]
%\begin{figure}
\includegraphics[height=6cm]{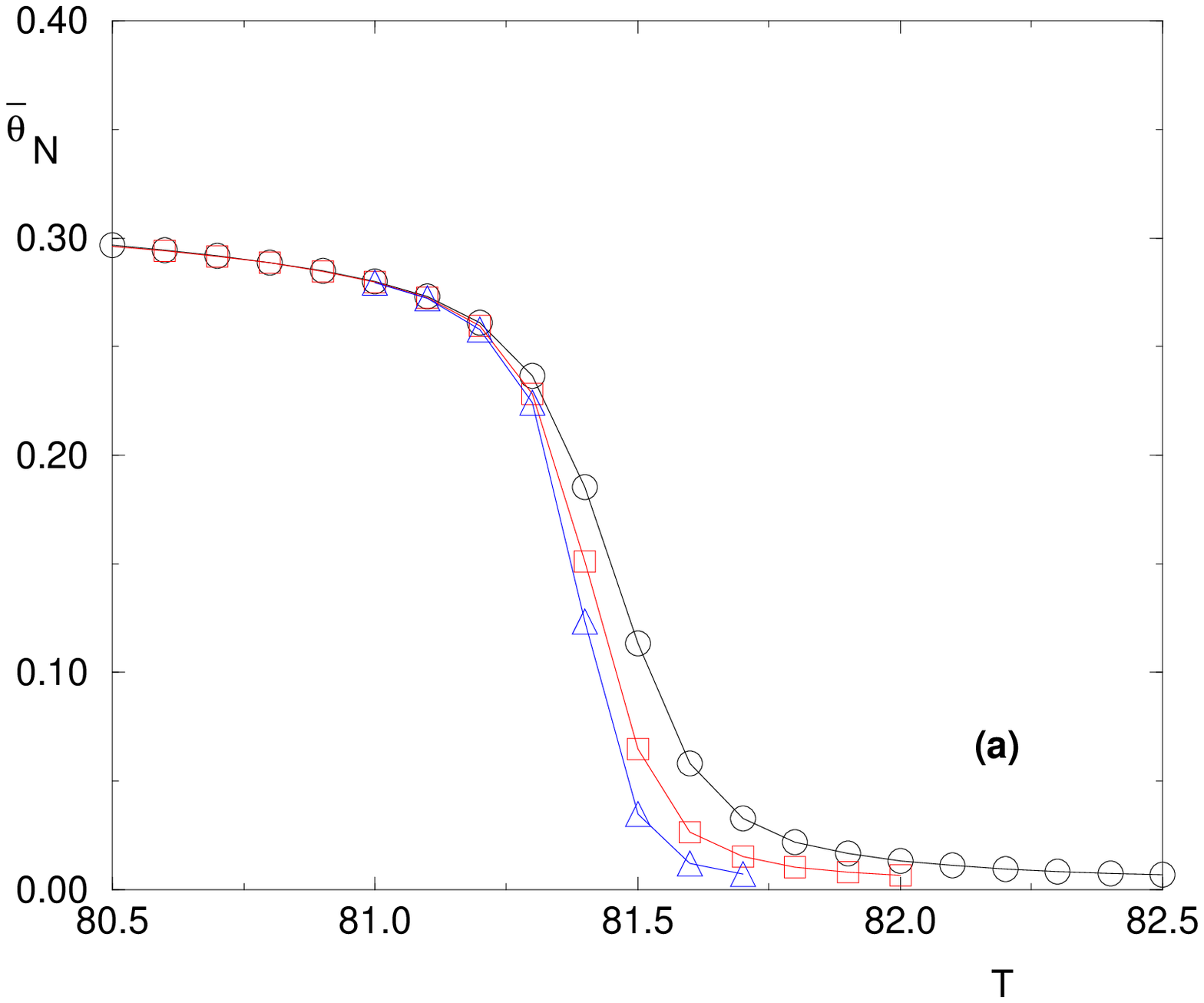}
\hspace{1cm}
\includegraphics[height=6cm]{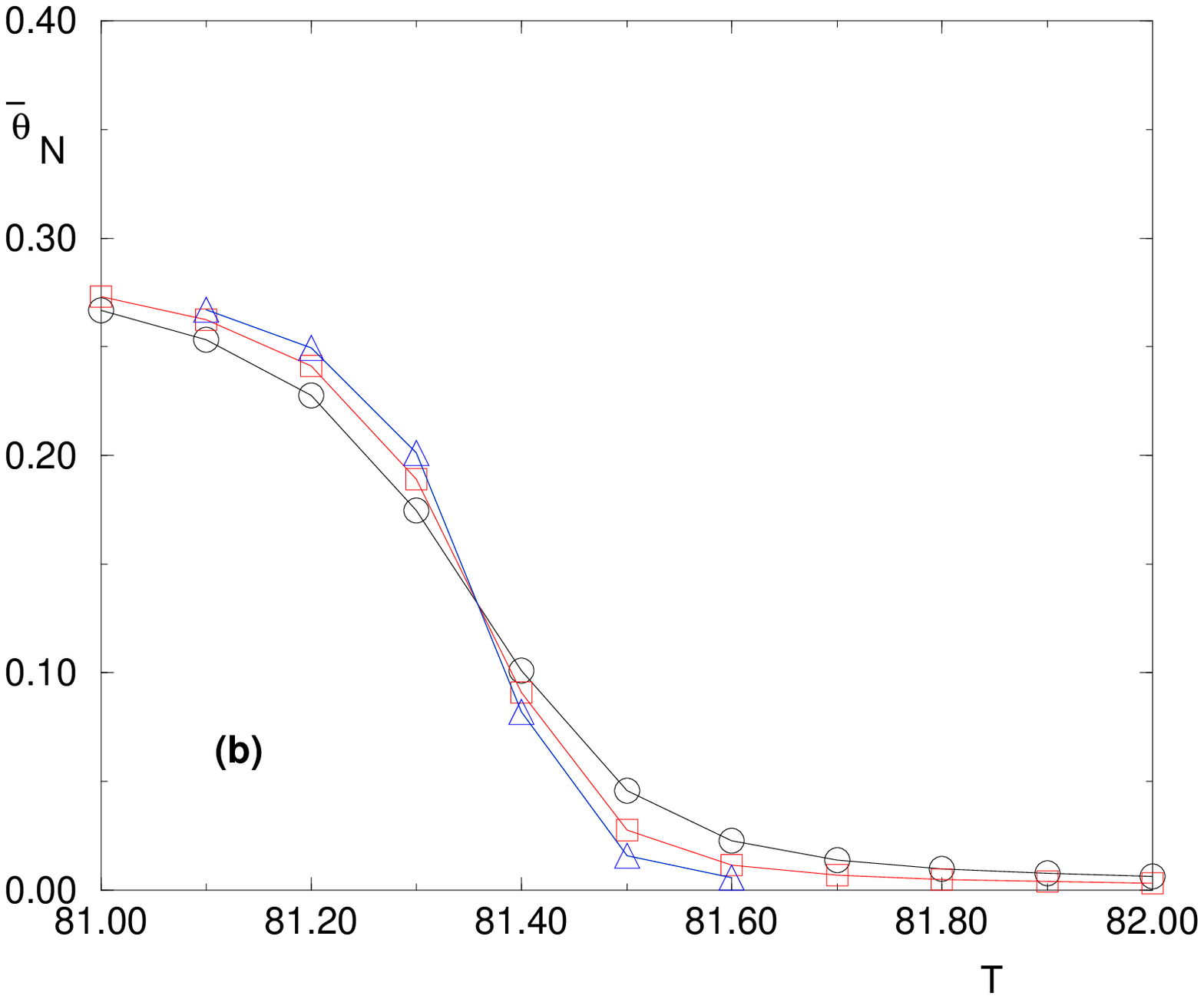}
\caption{The fraction of paired monomers $\overline{\theta}_N(T)$ of the
disordered 
case  for sizes $N=10^5$ $(\bigcirc)$, $2 \cdot 10^5$ $(\square)$, $4
\cdot 10^5$ $(\triangle)$, averaged over $10^4$ samples. The boundary
conditions are (a) bound-bound (b) bound-unbound. The error bars are
much smaller than the symbols.}
\label{f6}
\end{figure}

\begin{figure}[htbp]
%\begin{figure}
\includegraphics[height=6cm]{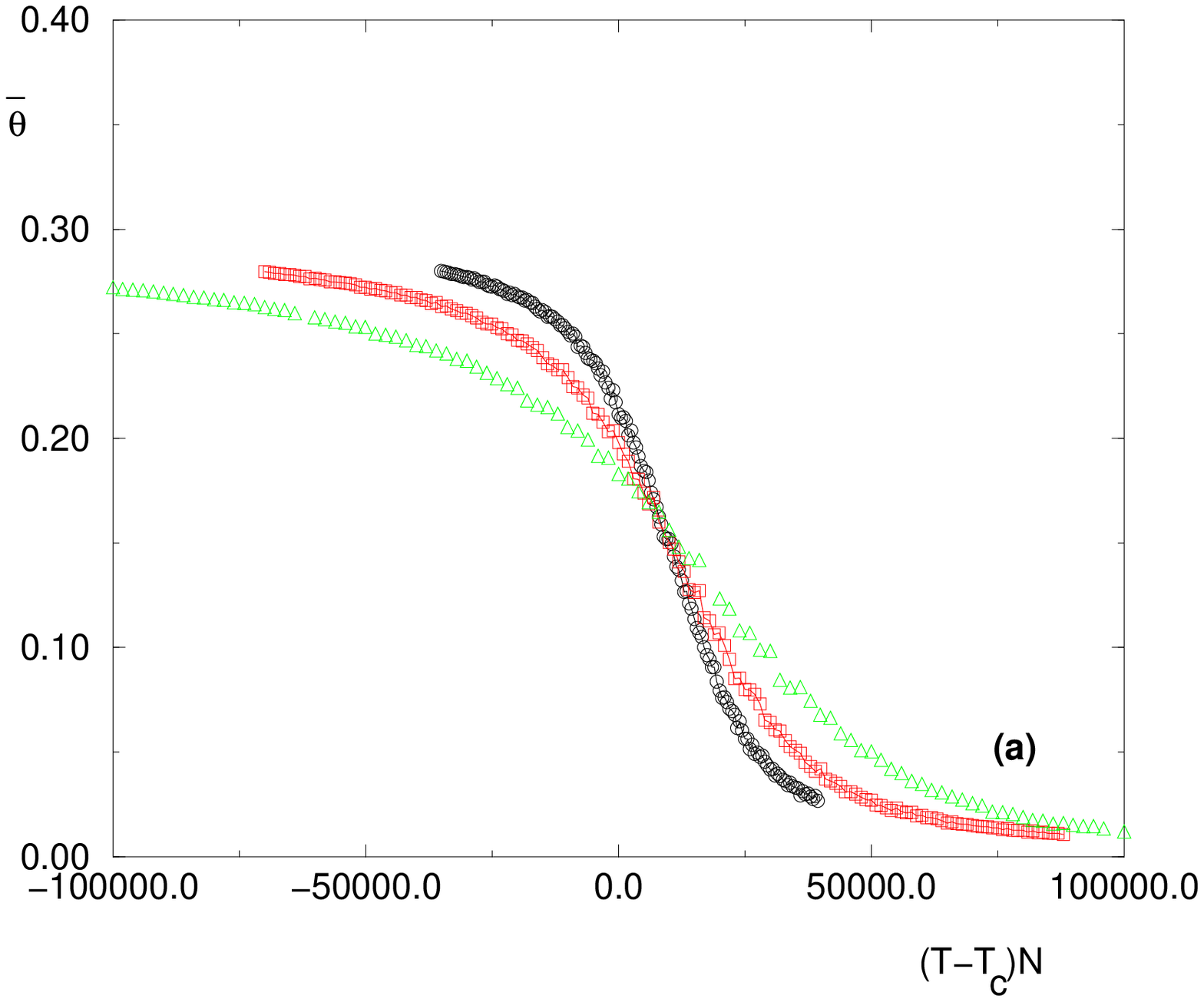}
\hspace{1cm}
\includegraphics[height=6cm]{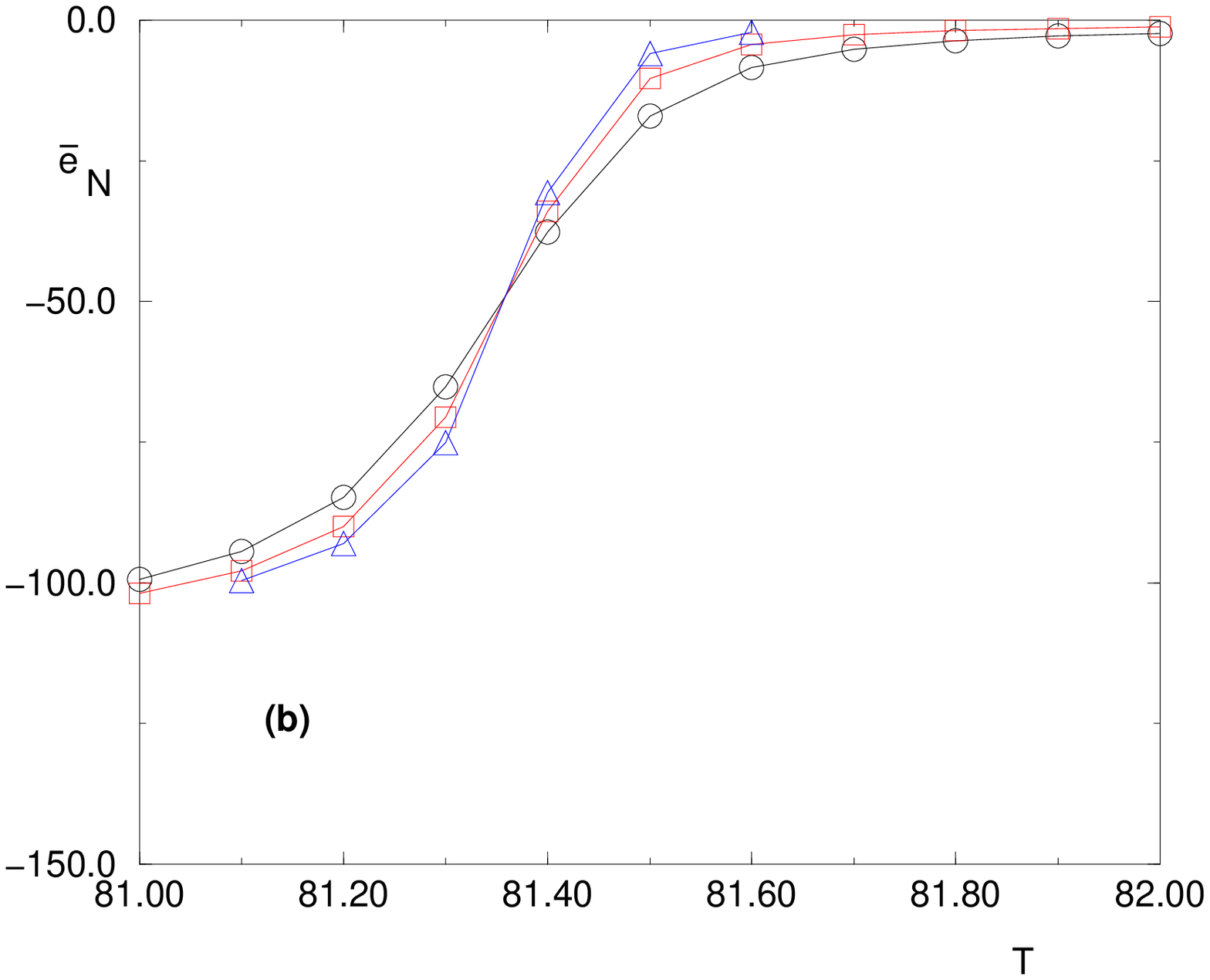}
\caption{(a) Tentative rescaling $\overline{\theta}_N(T)=\Theta(
(T-T_c)N )$ for the fraction of paired monomers, averaged over $500$ samples,
for sizes $N=10^5$ $(\bigcirc)$, $2 \cdot 10^5$ $(\square)$, $4 \cdot
10^5$ $(\triangle)$. The boundary condition is bound-bound (b) The
contact energy $\overline{e}_N(T)$ of the disordered case  for sizes
$N=10^5$ $(\bigcirc)$, $2 \cdot 10^5$ $(\square)$, $4 \cdot 10^5$
$(\triangle)$, averaged over $10^4$ samples. The boundary condition is
bound-unbound. The error bars are much smaller than the symbols.}
\label{f5}
\end{figure}

We have plotted on Fig. \ref{f6} the disorder averaged contact density
$\overline{\theta}_N(T)$ for the two boundary conditions :
the results are very similar to
the pure case results of Fig. \ref{f1}.
In particular, for the (bu) case, the crossing of the contact density
for different values of $N$ gives a direct determination of $T_c$, and
shows that the contact density is finite at criticality: the
transition is therefore first order. We also show in Fig. \ref{f5} (b) the
contact energy of eq. (\ref{energy}) for the (bu) case, which is
somewhat similar to the corresponding result for SAW's (see Fig 4 of 
ref \cite{Barbara2}). The crossing is again a clear sign of a discontinuous
transition.

However, in contrast with the pure case, the rescaling (\ref{rescalingtheta})
shown on Fig. \ref{f5} (a), is not satisfied at all
\begin{equation}
\overline{\theta}_N(T) \neq \Theta( (T-T_c)N ) 
\label{rescalingovertheta}
\end{equation}
Since the crossing of the contact density of Fig. \ref{f6} (b) 
implies that the transition is first-order,
we are led to the conclusion that there is a problem
with the finite-size scaling form 
for the disorder averaged contact density (\ref{rescalingovertheta}).
To understand why, we have studied the 
sample-to-sample fluctuations.

%\newpage

\subsection{ Sample to sample fluctuations }

\begin{figure}[htbp]
%\begin{figure}
\includegraphics[height=6cm]{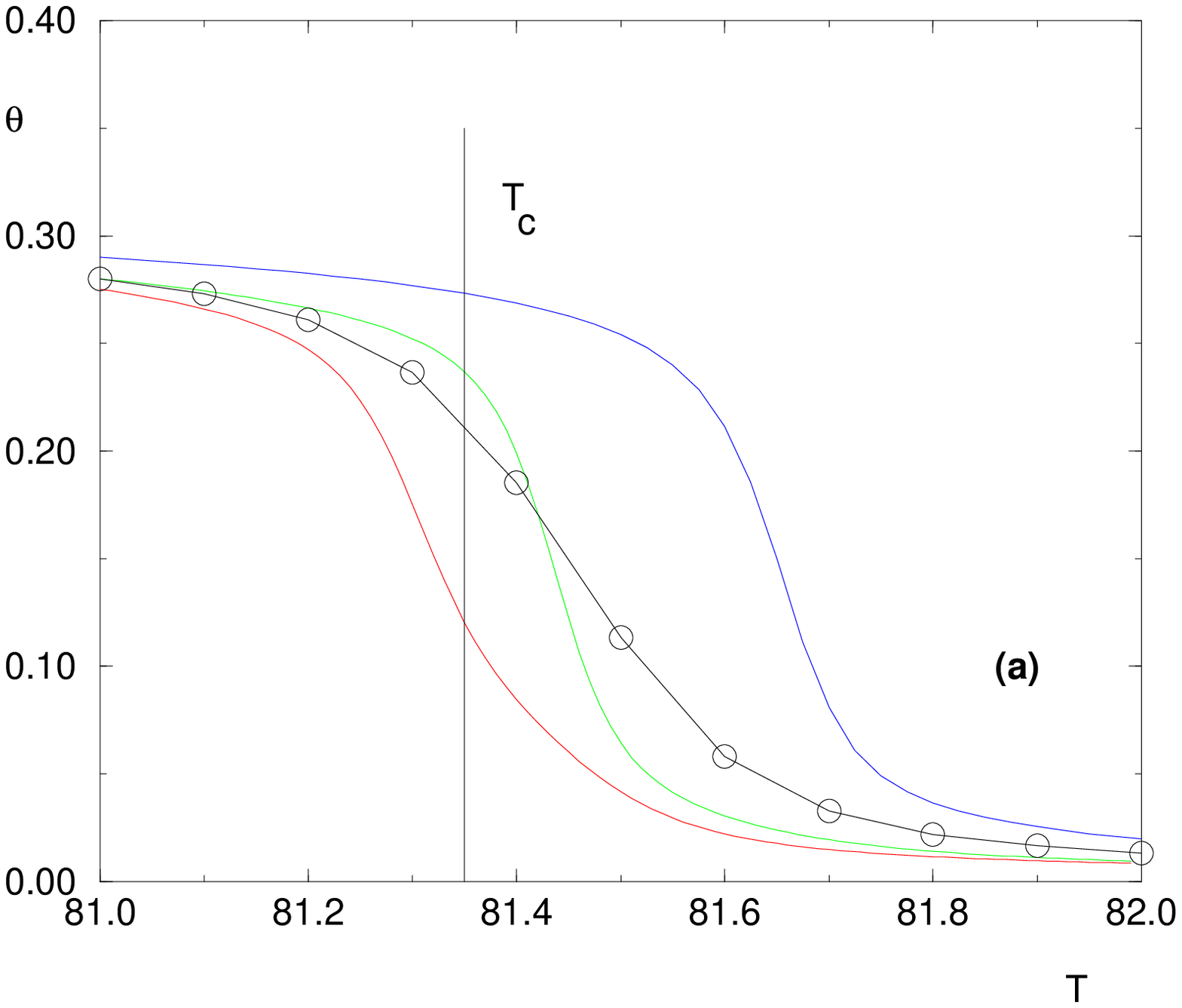}
\hspace{1cm}
\includegraphics[height=6cm]{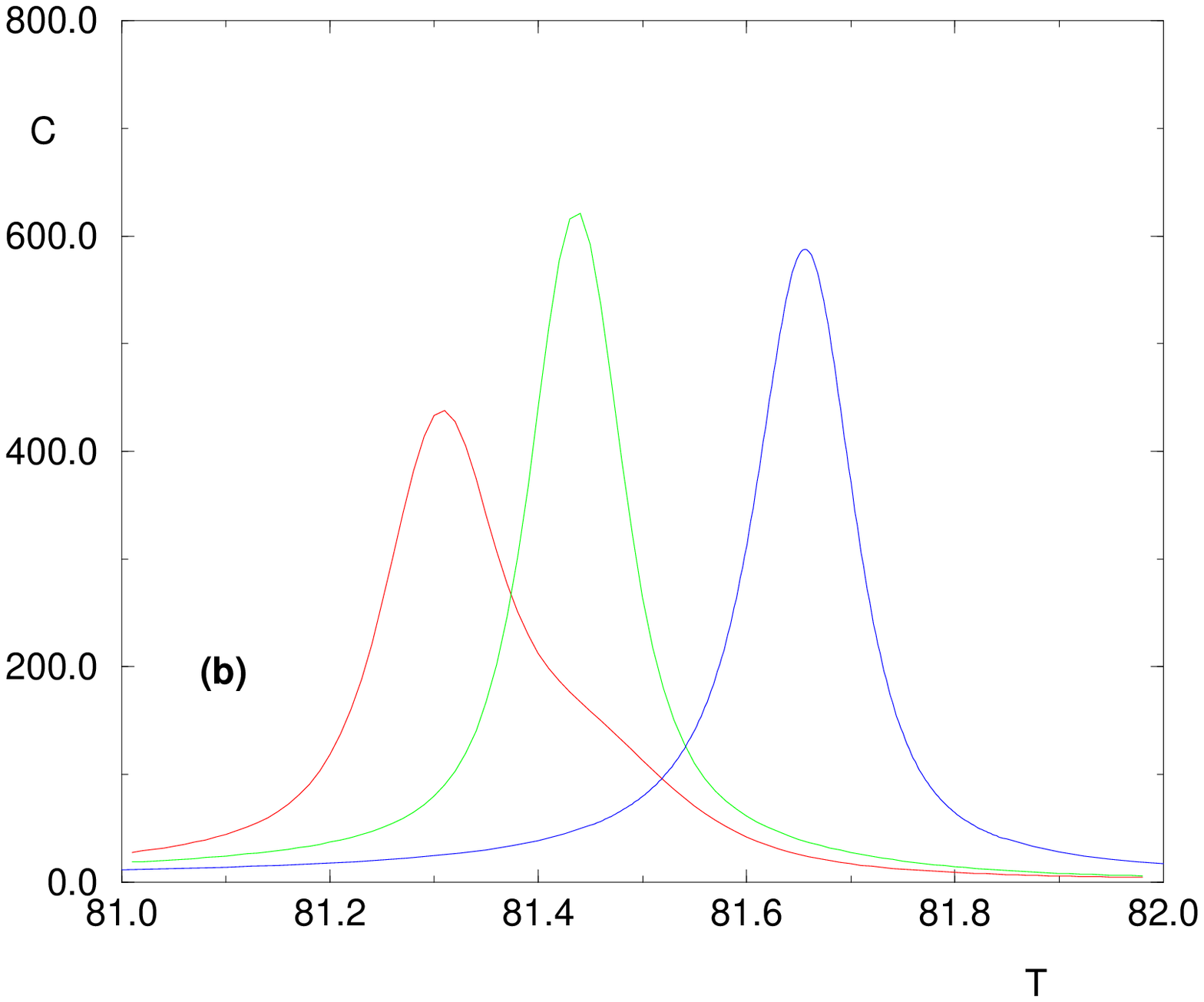}
\caption{(a) Comparison of the fraction of paired monomers of three
individual disorder samples with the average over $10^4$ samples
$(\bigcirc)$, for size $N=10^5$ and bound-bound boundary
condition. The thermodynamic critical temperature $T_c$ is shown (b)
The specific heat of the three disorder samples. }
\label{f8}
\end{figure}
On Fig. \ref{f8} (a), we have plotted the contact density
for three particular samples, as compared to 
the average over $10^4$ samples.
The sample to sample fluctuations are still strong
even for the size $N=10^5$.
In particular, at the critical temperature $T_c$ obtained from the crossing
of Fig. \ref{f6} (b), some samples are clearly already `delocalized',
whereas other remain localized up to a higher temperature.
On Fig. \ref{f8} (b), we have plotted the specific heat
for the three same samples : the peaks are clearly
separated.

These results are reminiscent of the lack of self-averaging
found in second order random critical points \cite{domany95,domany}.
In these systems, the argument goes as follows \cite{domany}
(i) off-criticality, the finite correlation length $\xi$
allows for a division of the sample into independent large
sub-samples, which in turn leads to the self-averaging property of
thermodynamic quantities (here, ``independent'' means that that the
interaction term between these sub-samples is a surface term, that can
be neglected with respect to volume contributions) (ii) at
criticality however, the previous 'subdivision' argument breaks down
because of the divergence of $\xi$. 

In usual first-order transitions, there is no diverging correlation
length. The analysis of the disorder effects consists in dividing
the system into finite sub-systems (or phases), and in taking into
account the surface tension between phases \cite{Im_Wo}.
The first order transition of the PS model is very different
for two polymeric reasons. First, there is no surface tension
between the localized and delocalized phases, as explained 
in section \ref{coexistence} on the coexistence at the pure critical point.
Second, there exists here an infinite correlation length, 
as is very clear in the (bu) case where the end segment length
diverges as $1/(T_c-T)$ \cite{Ka_Mu_Pe2}. In the (bb) case, this
diverging correlation length appears in the loop length distribution 
\begin{equation}
P_T^{pure}(l) \simeq \frac{1}{l^c} \ e^{- \frac{l}{\xi(T)}} \ \ \ {\rm with}
 \ \ \xi(T) \opsimeq_{T \to T_c^-} \frac{1}{T_c-T}  \
\end{equation}

We are therefore led to the conclusion that
the lack of self-averaging we find in the disordered PS model
comes from the presence of an infinite correlation length,
(even though the transition is first order), that prevents
the division of the chain into independent sub-chains.

The strong sample-to sample fluctuations explain
why the finite-size scaling analysis (\ref{rescalingovertheta})
does not work for the disordered averaged contact density :
each sample $(i)$ has its own pseudo-critical temperature $T_c(i)$
where it presents a pseudo-first order transition,
and the resulting disorder average has no nice
finite-size scaling behavior.

\subsection{ Finite-size scaling for the contact density in a given
sample}

Since we have obtained that there are strong sample to sample
fluctuations that invalidates the usual finite size scaling analysis
for the disorder averaged contact density,
we have tried to use more refined finite size scaling theories
in the presence of disorder \cite{domany95,Paz1,domany,Paz2} :
in these references, it is stressed that the
sample to sample fluctuations of a given observable
are predominantly due to the sample to sample fluctuations
of the pseudo-critical temperatures $T_c(i,N)$.
As a consequence, these references express the finite size
scaling form of an observable $X$ \cite{domany,Paz2} as
\begin{equation}
X_N^{(i)}(T) = N^{\rho} Q_i \left( (T-T_c(i,N)) N^{\phi} \right)
\label{fssdomany}
\end{equation}
where $T_c(i,N)$ is the pseudo-critical temperature
of the sample $i$, determined for instance
as the susceptibility peak \cite{domany},
or by a minimal distance criterium \cite{Paz2}.
The scaling function $Q_i$ a priori also depends
on the sample, and we refer to \cite{domany,Paz2}
for discussions and examples.

Here, from the specific heat plotted on Fig \ref{f8}b,
we see that beyond the
sample dependence of the pseudo-critical temperature $T_c(i)$,
there is also a sample dependent scaling function $Q_i$,
in contrast with the case studied in \cite{Paz2},
where the curves of various samples were the same up
to a translation $T-T_c(i)$.

\begin{figure}[htbp]
%\begin{figure}
\includegraphics[height=6cm]{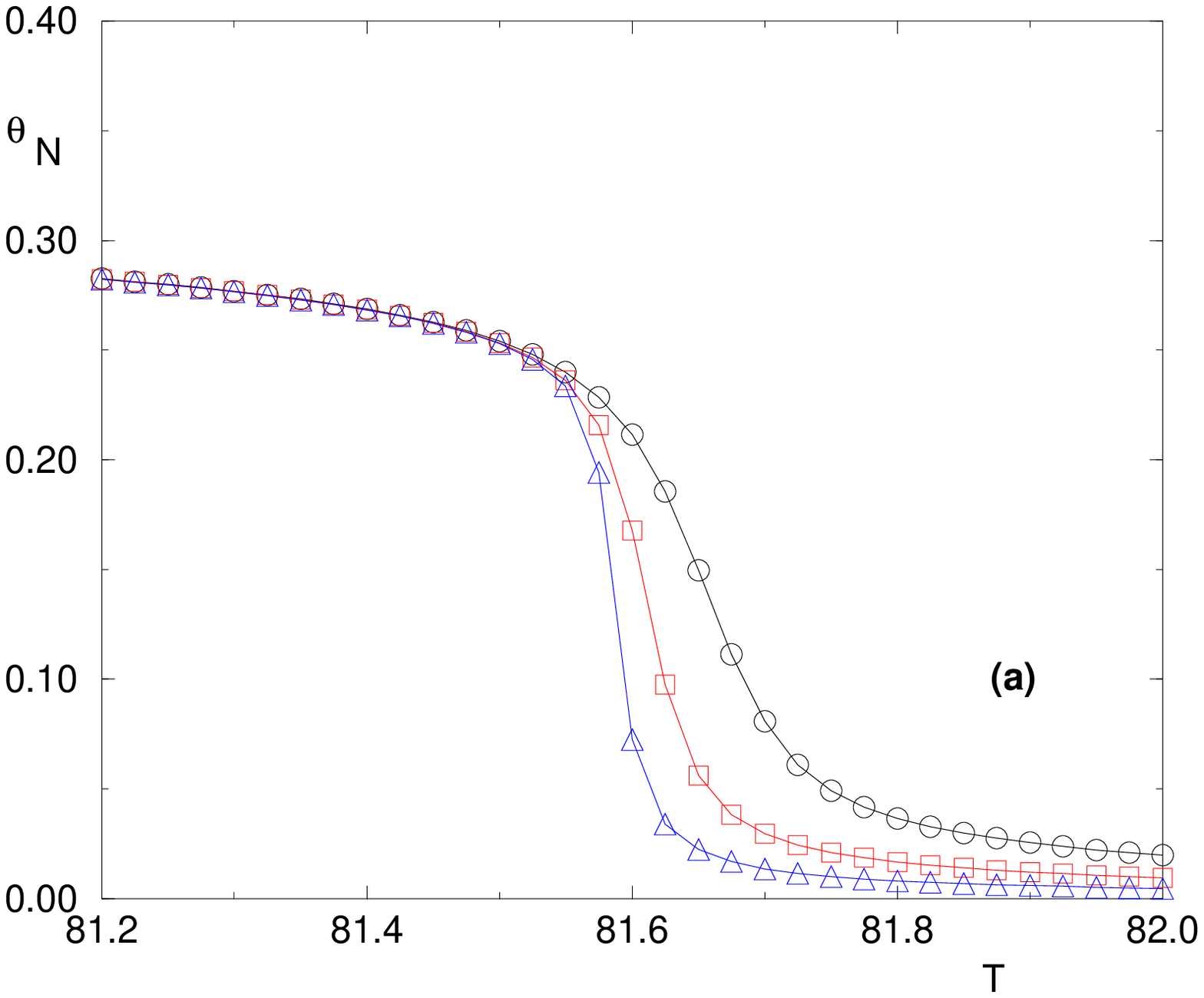}
\hspace{1cm}
\includegraphics[height=6cm]{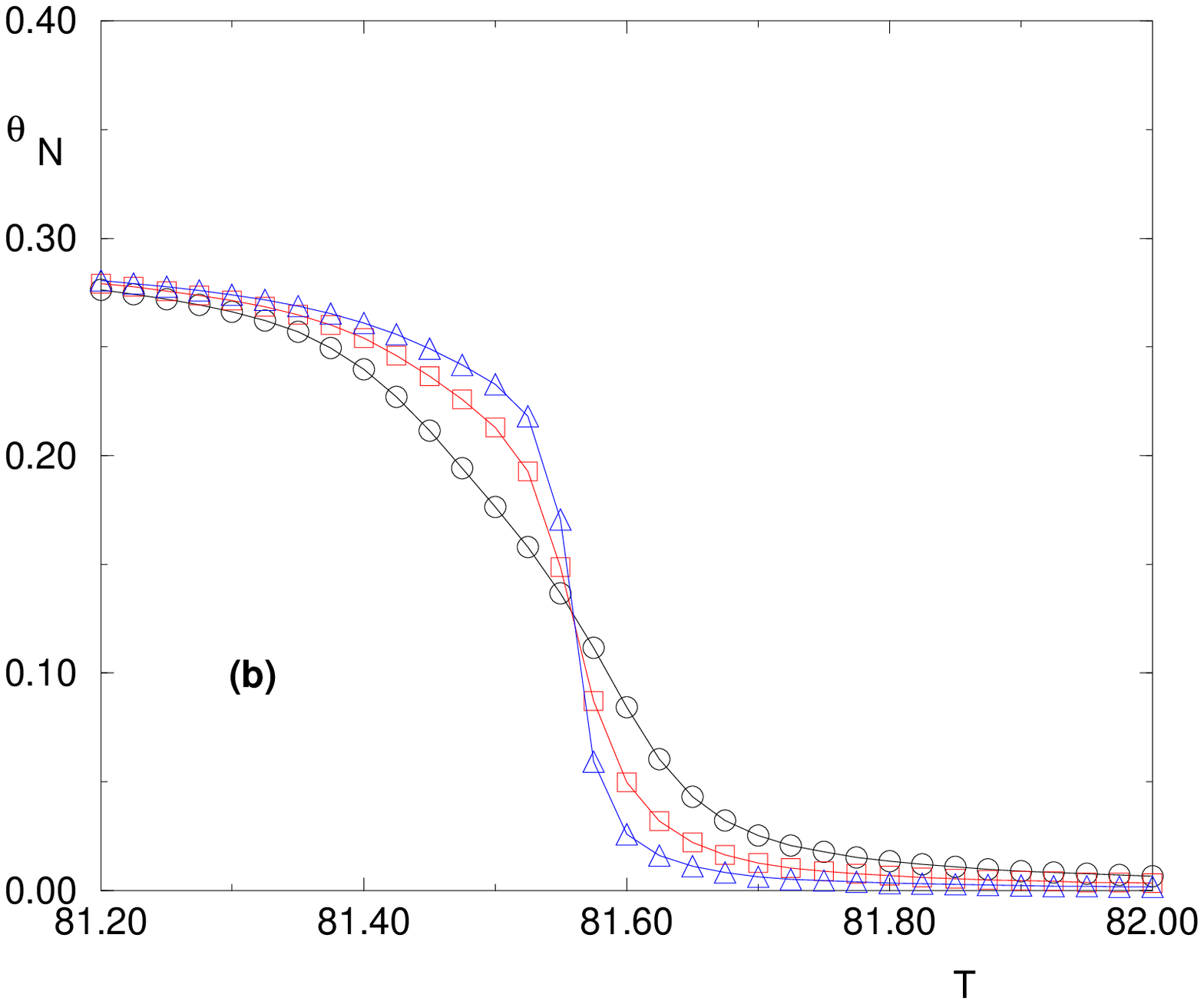}
\caption{The fraction of paired monomers $\theta_N(T)$ for a single
disorder sample, using the 'sample replication' procedure discussed in 
the text, and for sizes $N=10^5$ $(\bigcirc)$, $2 \cdot 10^5$
$(\square)$, $4 \cdot 10^5$ $(\triangle)$. The boundary conditions are
(a) bound-bound (b) bound-unbound.} 
\label{f9}
\end{figure}

\begin{figure}[htbp]
%\begin{figure}
\includegraphics[height=6cm]{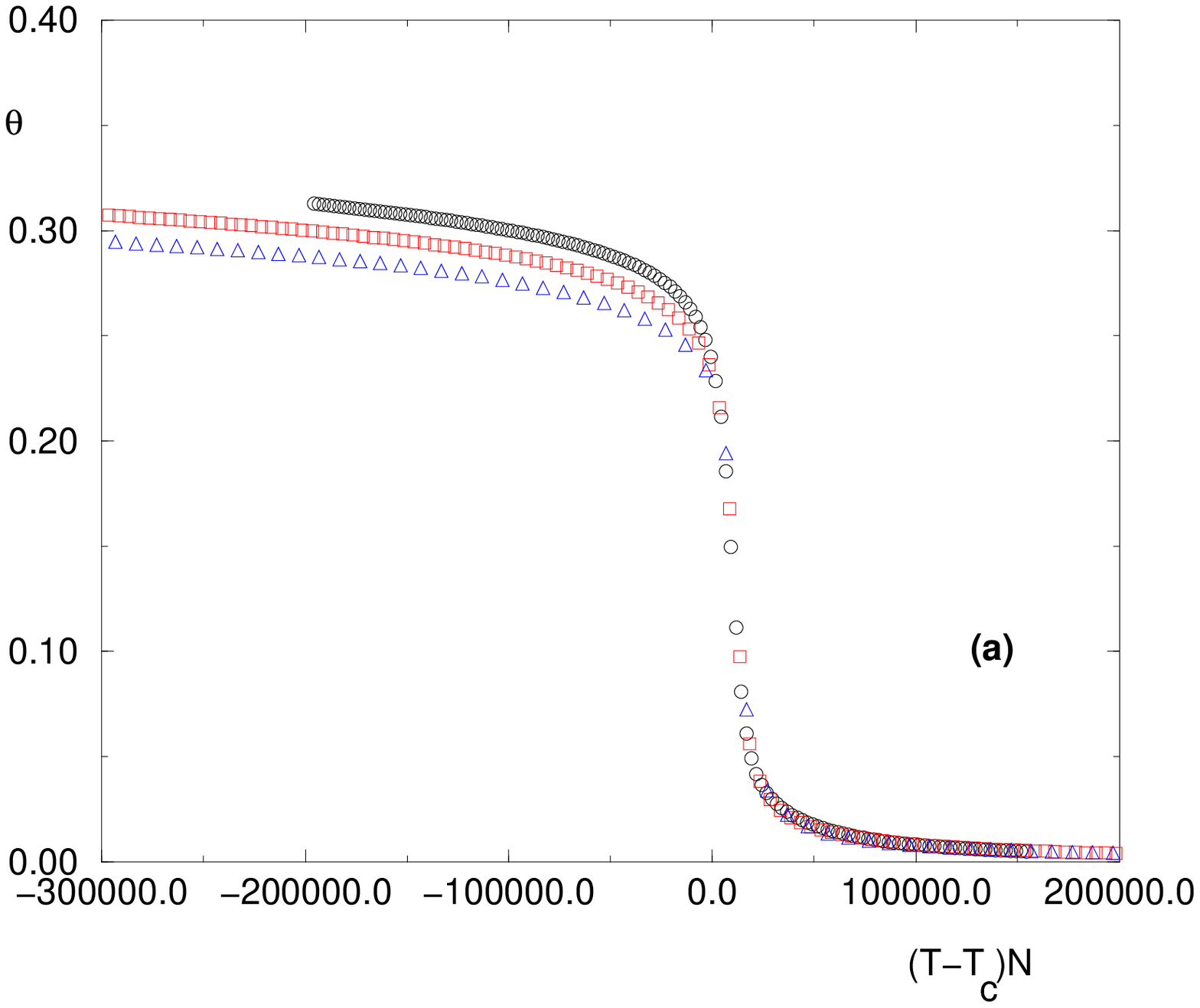}
\hspace{1cm}
\includegraphics[height=6cm]{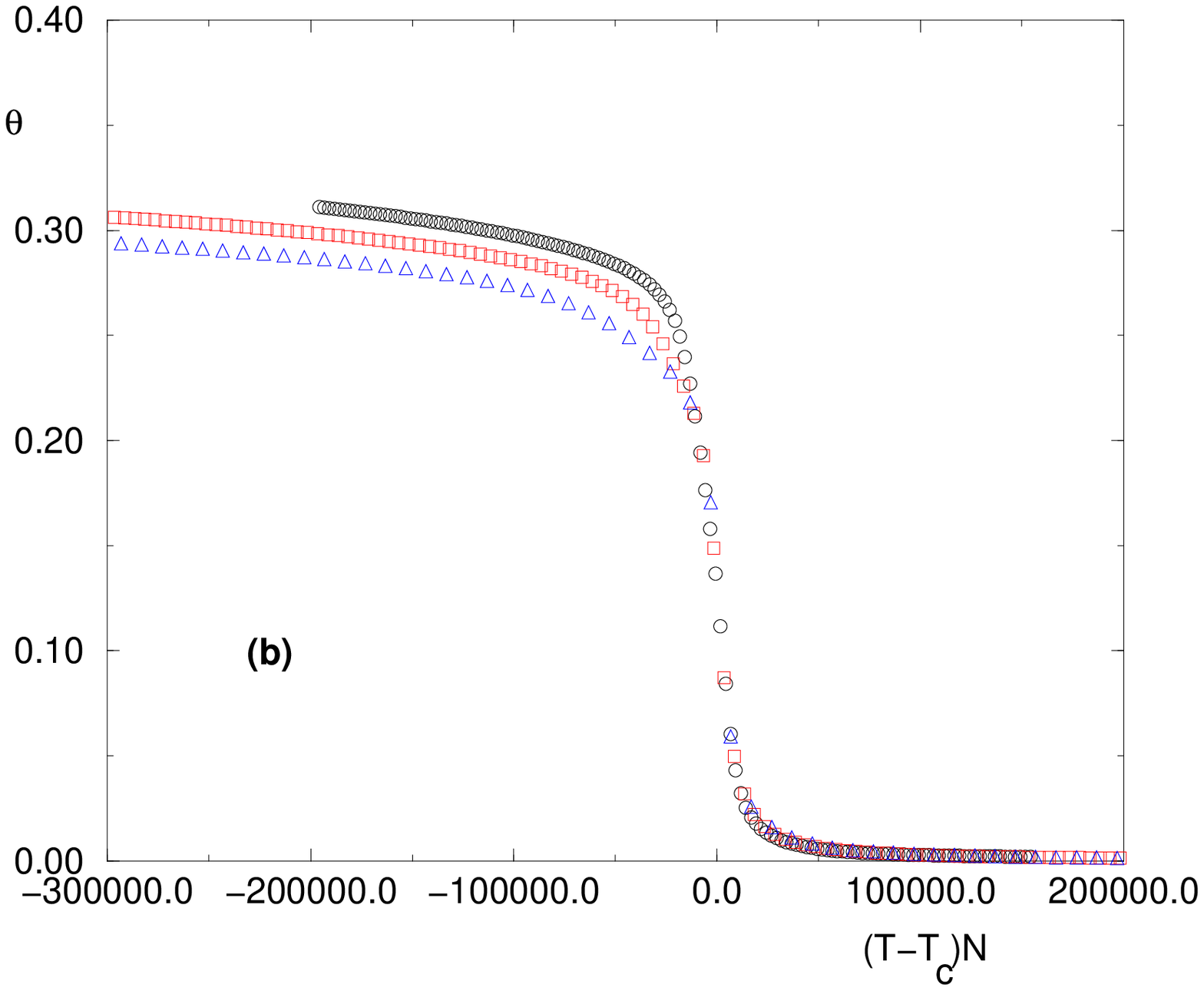}
\caption{The scaled fraction of paired monomers
$\theta((T-T_c)N))$ for the same disorder sample as in Fig
\ref{f9}, for sizes $N=10^5$ $(\bigcirc)$, $2 \cdot 10^5$ $(\square)$, $4 \cdot
10^5$ $(\triangle)$. The boundary conditions are (a) bound-bound (b)
bound-unbound.} 
\label{f10}
\end{figure}

\begin{figure}[htbp]
%\begin{figure}
\includegraphics[height=6cm]{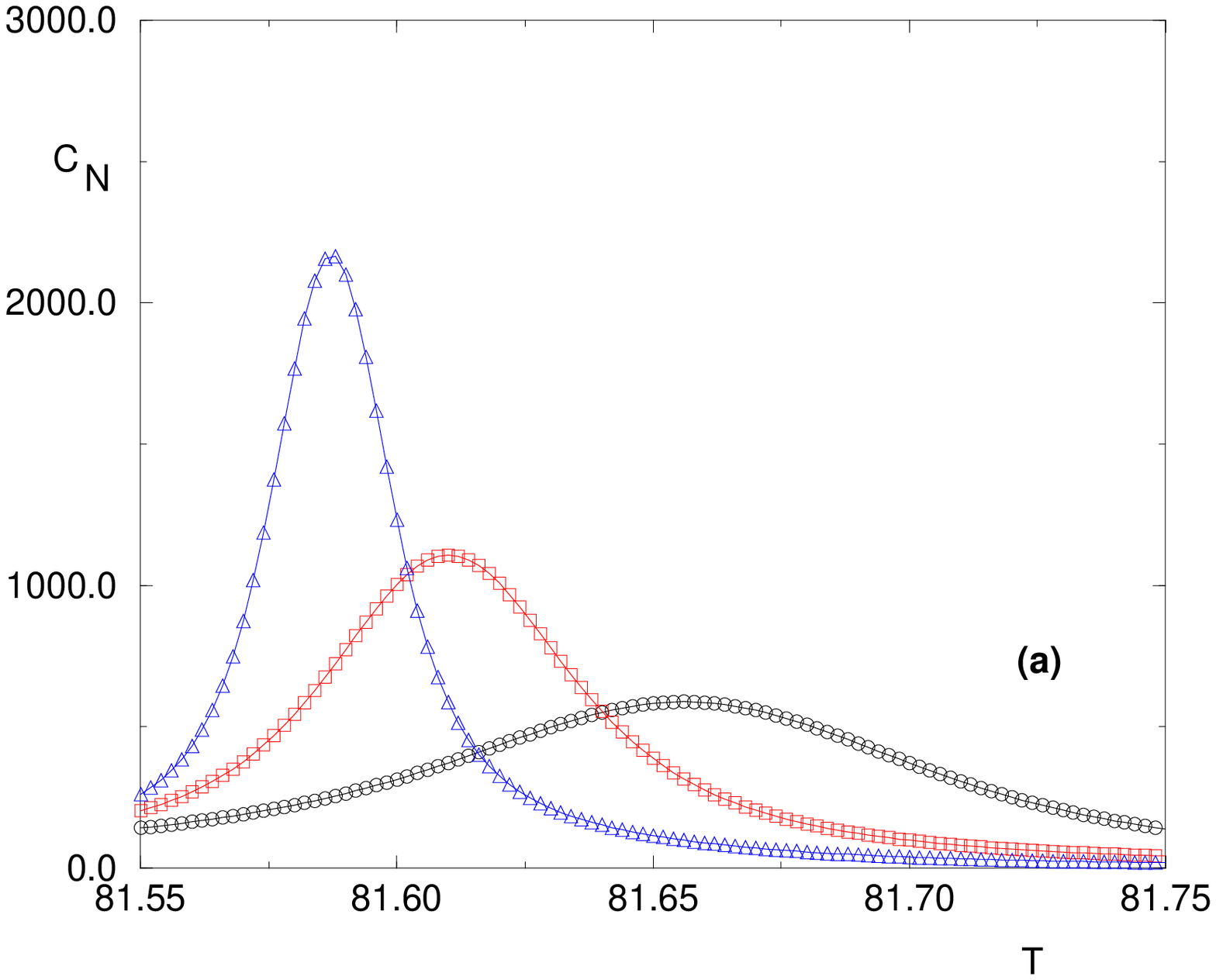}
\hspace{1cm}
\includegraphics[height=6cm]{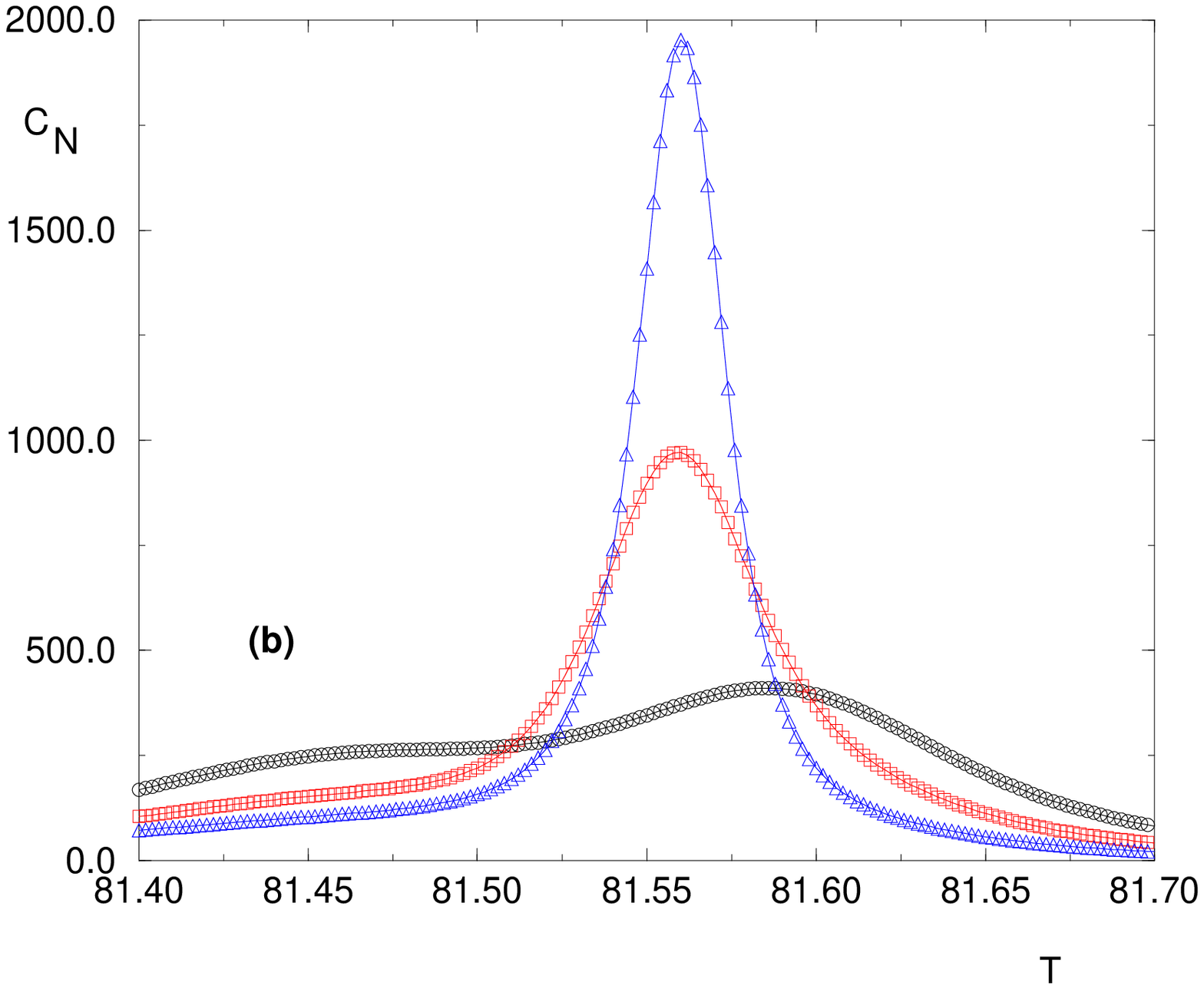}
\caption{The specific heat $C_N$ for the same disorder sample as in Fig
\ref{f9} and Fig \ref{f10}, for sizes $N=10^5$ $(\bigcirc)$, $2 \cdot
10^5$ $(\square)$, $4 
\cdot 10^5$ $(\triangle)$. The boundary conditions are (a) bound-bound
(b) bound-unbound.} 
\label{f11}
\end{figure}

As a consequence, we have tried to obtain some information
from a sample-dependent finite-size scaling analysis
via the following 'sample replication' procedure :
we have compared a given random sample $E=(\epsilon_1,...,\epsilon_N)$ of
size $N=10^5$
with the sample of size $2N=2 . 10^5$ obtained by gluing together two copies
of the initial sample $E$,
and with the sample of size $4N=4 . 10^5$ obtained by gluing together four
copies of the initial sample $E$ : the corresponding results for the
contact density
for the two boundary conditions are shown on Fig. \ref{f9}.
In particular, the crossing obtained for the (bu) boundary conditions
allows to determine the sample-dependent pseudo-critical temperature
$T_c(i)$.
The rescaling of these data with the reduced variable $x=(T-T_c(i))N$
is found to be satisfactory in the delocalized phase $T>T_c(i)$
for the two boundary conditions, see Fig. \ref{f10}.
We have also plotted on Fig. \ref{f11} the associated specific heats :
for the two boundary conditions, the maximum value scales as $N$.

\newpage

\subsection{ Pairing probability along the chain in a given sample }
\begin{figure}[htbp]
%\begin{figure}
\includegraphics[height=6cm]{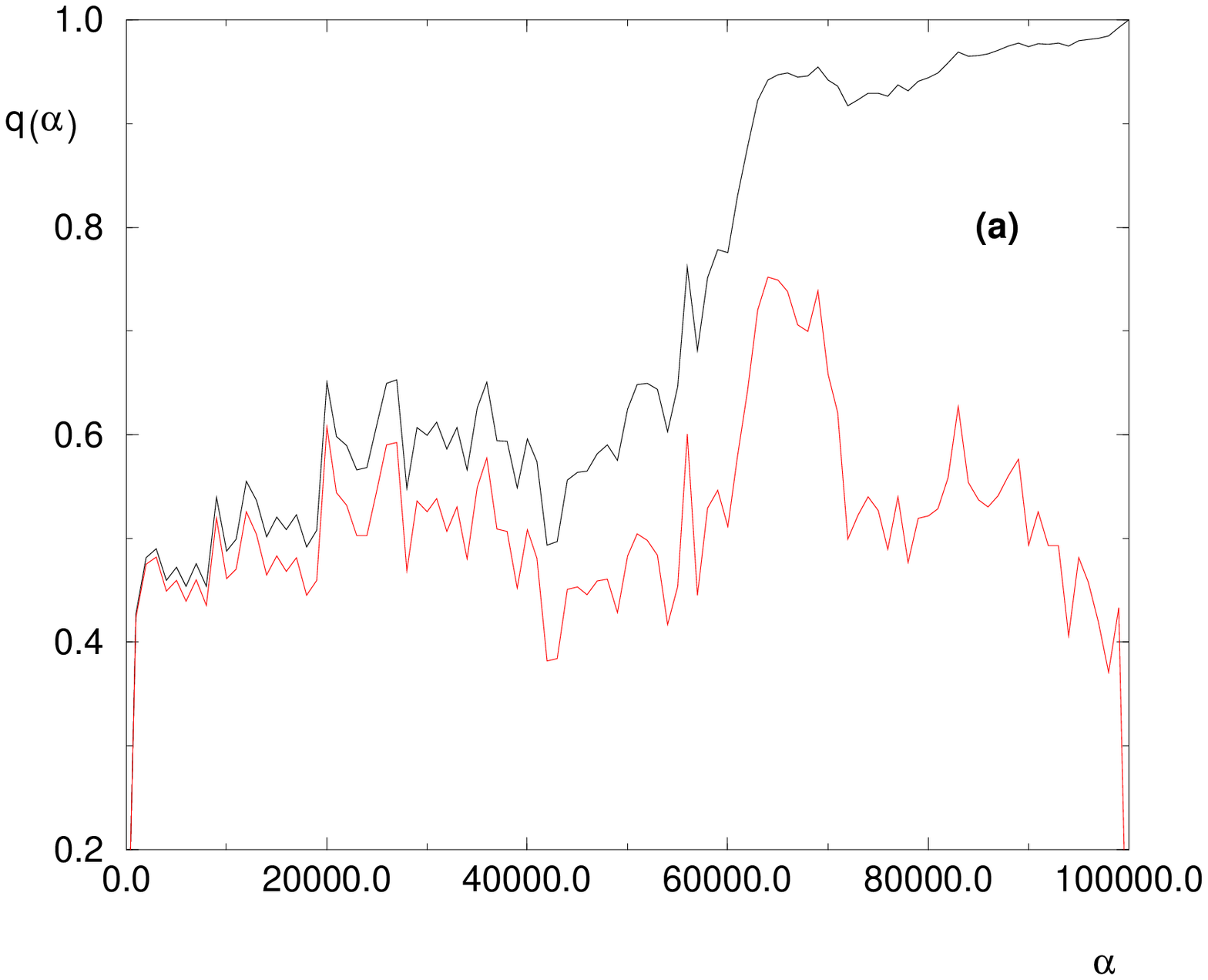}
\hspace{1cm}
\includegraphics[height=6cm]{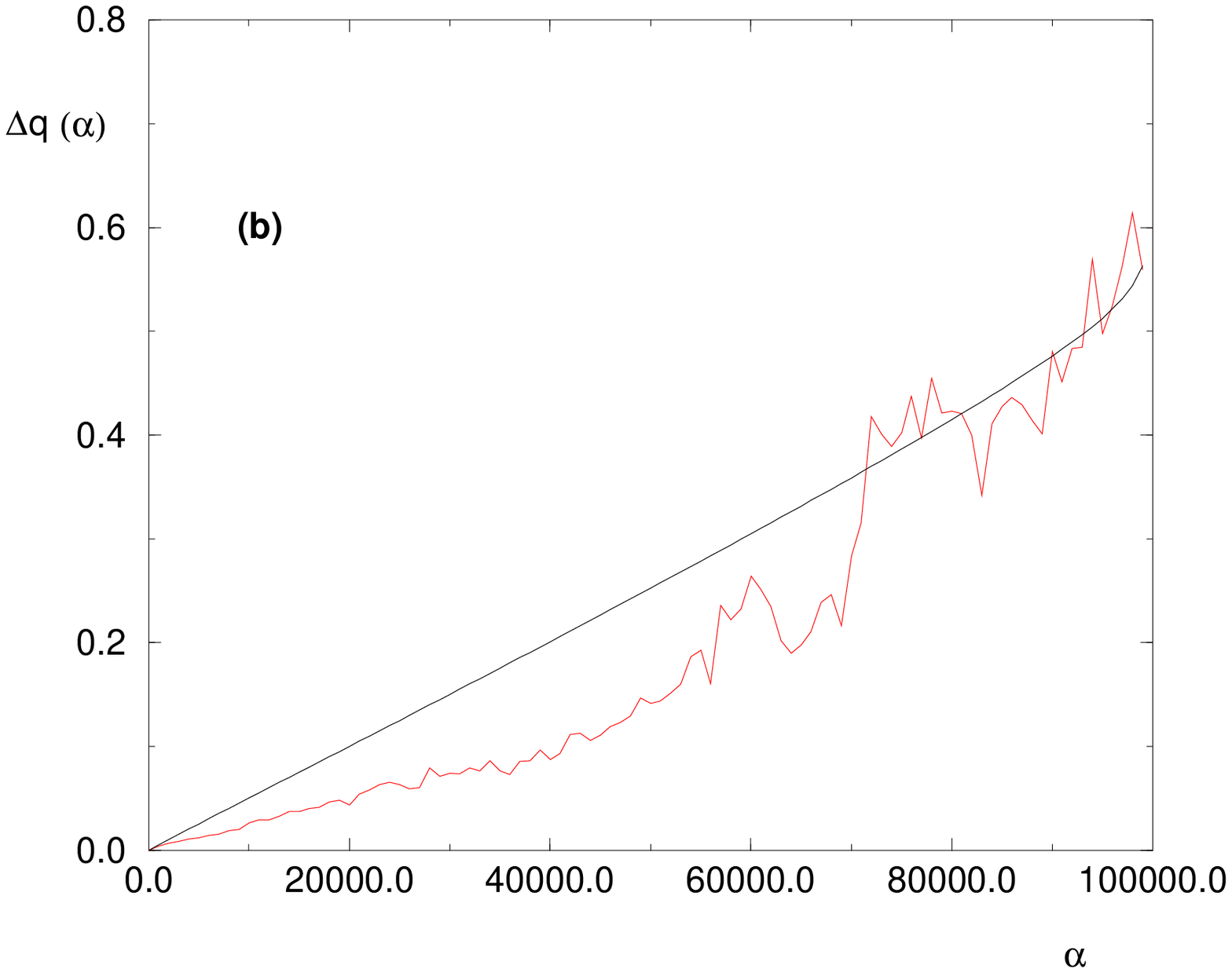}
\caption{(a) The unpairing probability $q(\alpha)$ for a
single disorder sample $(i)$ at its critical temperature $T_c(i)$, for
size $N=10^5$. The boundary conditions are bound-bound (lower curve)
and bound-unbound (upper curve)(b) The difference $\Delta
q(\alpha)=q_{bu}(\alpha)-q_{bb}(\alpha)$
between the two previous curves compared with the analog quantity of
the pure case at $T_c$ (see Figure \ref{f4} (a)).}
\label{f12}
\end{figure}

On Fig. \ref{f12}(a), we have plotted, for a given sample at its
pseudo-critical
temperature $T_c(i)$ determined by Fig. \ref{f9} (b), the unpairing
probability $q(\alpha)$ as a function of the monomer index
$\alpha$
for the two boundary conditions. These data should be compared
with the pure case results of Fig. \ref{f4} : beside the disorder
fluctuations present in the (bb) boundary conditions, the (bu) boundary
conditions display an upward trend that shows that there exists some
phase coexistence between a localized and delocalized phase. To show
more clearly this coexistence, we have plotted  in Fig.  \ref{f12}(b)
the difference $\Delta q(\alpha)=q_{bu}(\alpha)-q_{bb}(\alpha)$ for
the same disordered sample, as compared to the linear analog of the
pure case. 

\newpage

\subsection{ Disorder averaged loop statistics}
\begin{figure}[htbp]
%\begin{figure}
\includegraphics[height=6cm]{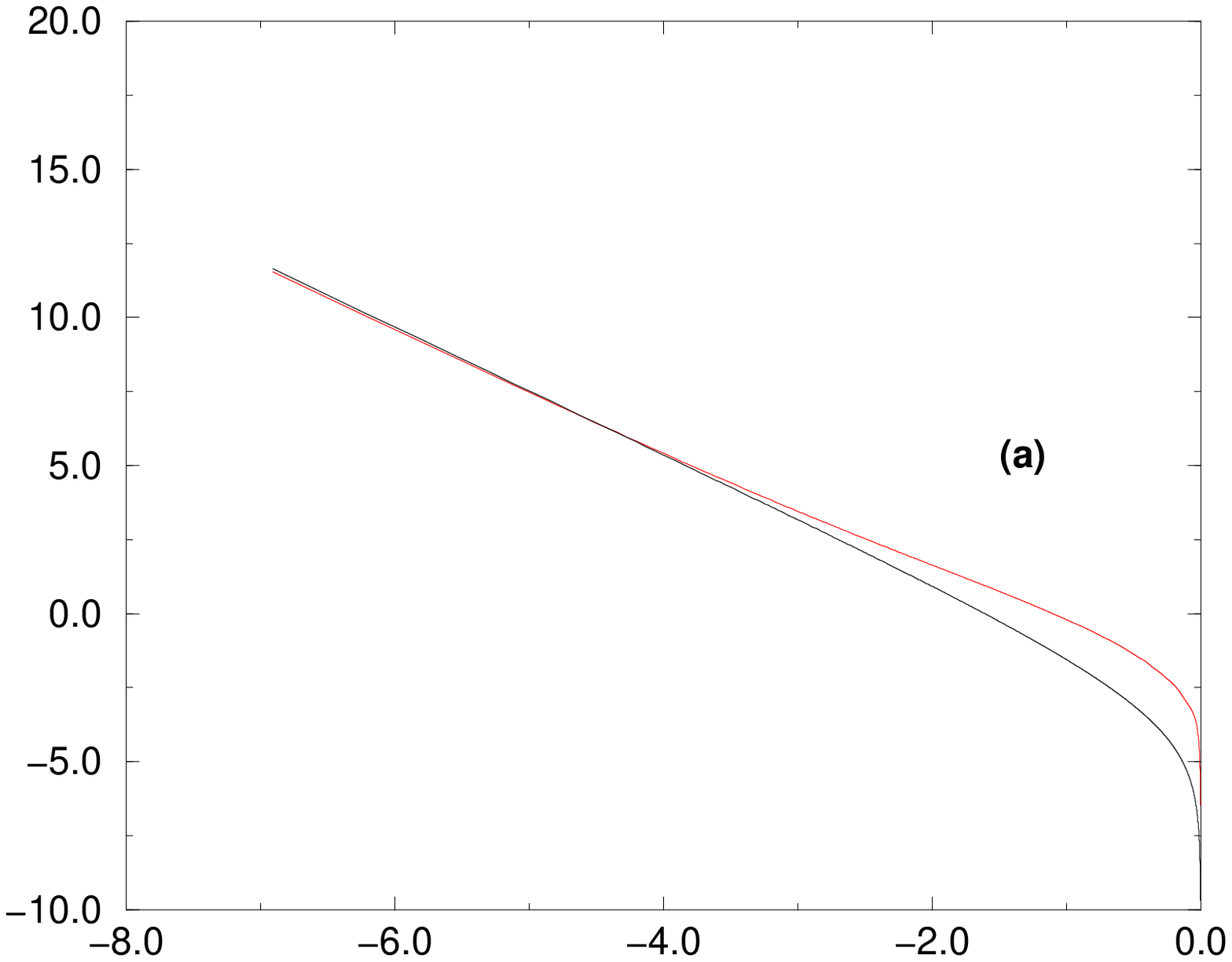}
\hspace{1cm}
\includegraphics[height=6cm]{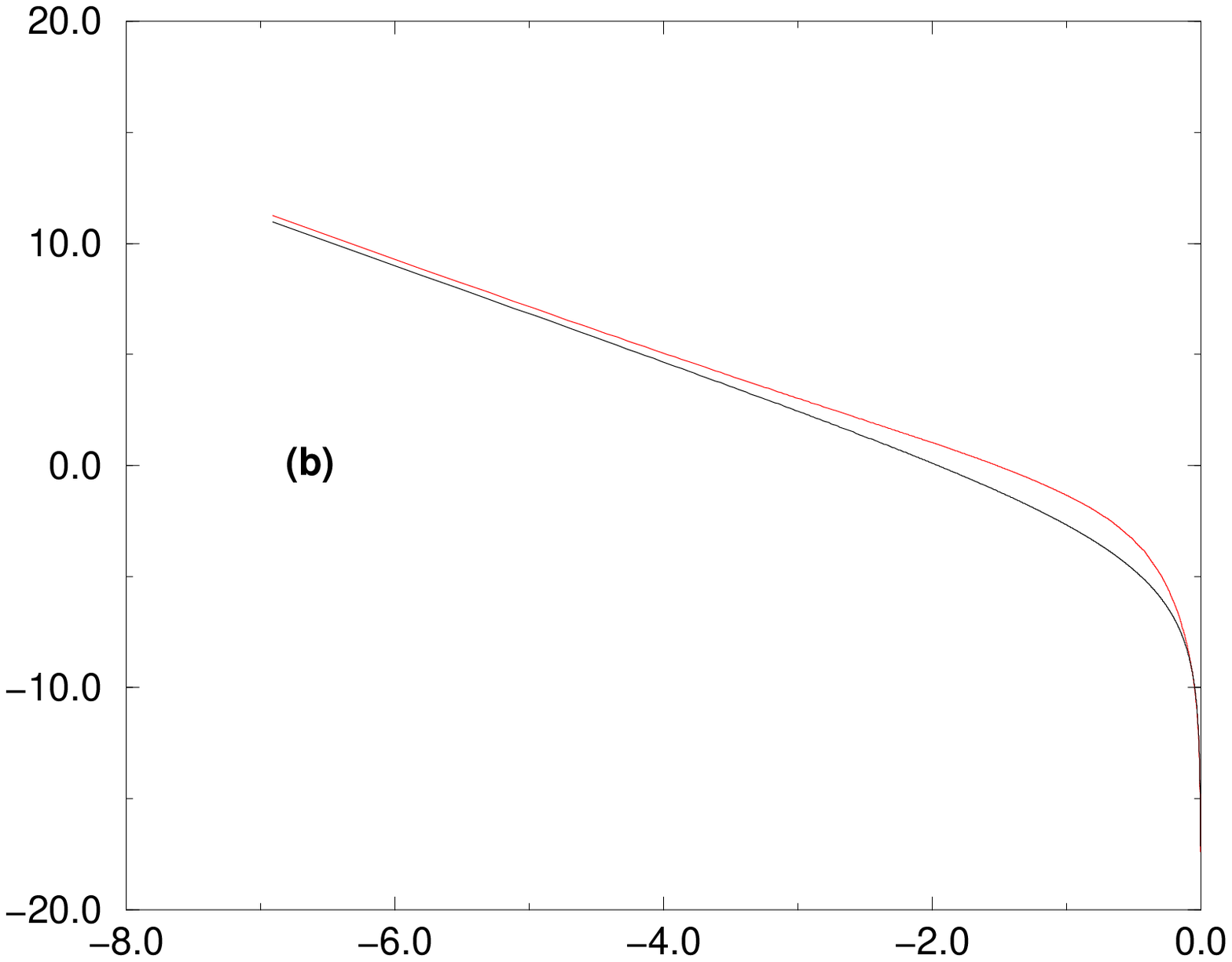}
\caption{ Log-Log plot of the critical loop distribution
$M_N(l)$ in the rescaled variable $\lambda =\frac{l}{N}$ for the
disordered case (averaged over $10^4$ samples) and for the pure case
(lower curve). Both systems have  
size $N=10^5$ and the boundary conditions are (a) bound-bound
(b) bound-unbound.}
\label{f13}
\end{figure}

We now turn to the probability measure $M_N(l)$
of the loops of length $l$ existing in a sample of size $N$,
as defined in eq. (\ref{pnl}). In Fig. \ref{f13}, we present the
disorder averaged $\overline{M_N(l)}$ at criticality,
in terms of the rescaled length $\lambda=l/N$ for the two boundary
conditions (bb) and (bu). In each case, we have added the corresponding
pure critical analog for comparison.

In the pure case, the loop length distribution at criticality
is given in the thermodynamic limit
by the loop weight used to defined the PS model (\ref{asymp})
$P(l) \sim 1/l^c$. Here in the pure finite sample, the exponent $c=2.15$
indeed corresponds to the slope in the regime $l \ll N$, before
the appearance of some curvature near $l \sim N$.
This form for the pure critical loop statistics is
thus very similar to the corresponding result for pure SAWs \cite{Carlon,
Baiesi1, Baiesi2}.
In the disordered case, we clearly obtain the same slope $c=2.15$
in the regime $l \ll N$, and deviations with respect to the pure case
only occur in the region $l \sim N$.
Our conclusion is thus that in the presence of disorder,
the statistics of loops at criticality remains $P(l) \sim 1/l^c$ in
the thermodynamic limit as in the pure case.

\section{Conclusion }

In this paper, we have studied the binary disordered Poland-Scheraga
model of DNA denaturation, in the regime where the pure model displays
a first order transition (loop exponent $c>2$). Our main conclusion is
that the transition remains first order in the disordered case, but
that disorder averaged observables do not satisfy finite size scaling,
as a consequence of strong sample to sample fluctuations.

Let us now consider the relation between the PS model and the Monte
Carlo simulations of attracting SAW's. In the pure case, the
finite-size behavior we have presented for the PS model 
with (bu) boundary conditions is very similar to the corresponding
behavior obtained previously in Monte-Carlo simulations of SAW's
\cite{Barbara1}. In the disordered case, our result for the crossing
of the contact energy (Fig \ref{f5} (b)) is somewhat similar to the
corresponding results  (Fig 4) of ref. \cite{Barbara2}. The failure of 
finite size scaling for disordered averaged quantities was also noted
in this reference. It therefore seems that the PS model with (bu)
boundary conditions is an appropriate effective model for SAW's even
in the presence of disorder.

{ \bf Acknowledgments : } 
We are grateful to B. Coluzzi for communicating
her results \cite{Barbara2} on disordered SAWs prior to publication: they
motivated the present study on the PS model. We thank Y. Kafri and
H. Orland for discussions.

\end{document}